%% file: paper.tex
\documentclass[journal]{IEEEtran}
\usepackage{ifpdf}
\usepackage{color}
\usepackage[latin1]{inputenc}
\usepackage{cite}
\usepackage{mathtools}
\usepackage{amssymb}
\usepackage{mathtools}
\usepackage{algorithm}
\usepackage{algpseudocode}
\usepackage{bigints}
\usepackage{url}
\usepackage{framed}

\usepackage{subfigure}

\ifpdf
\usepackage{graphicx} \DeclareGraphicsExtensions{.png} \graphicspath{{Figures}}
\else
\usepackage{graphicx} \DeclareGraphicsExtensions{.pdf} \graphicspath{{Figures}}
\usepackage{graphicx} \DeclareGraphicsExtensions{.eps} \graphicspath{{Figures}}
\fi

\usepackage{tikz}
\usetikzlibrary{decorations.pathmorphing}
\usetikzlibrary{fit}
\usetikzlibrary{backgrounds}
\usetikzlibrary{shapes}
\usetikzlibrary{positioning}
\usetikzlibrary{shadows}
\usetikzlibrary{decorations.pathmorphing}
\usetikzlibrary{decorations.shapes}
\usepackage{multirow}
\usepackage{dcolumn}
\usepackage{pgfplots}
\pgfplotsset{compat=1.3}

\begin{document}

\title{Cooperative Relaying under Spatially and Temporally Correlated Interference}
\author{Alessandro Crismani\thanks{Alessandro Crismani, Udo Schilcher, G\"{u}nther Brandner, and Christian Bettstetter are with the Mobile Systems Group, Institute of Networked and Embedded Systems, University of Klagenfurt, Klagenfurt 9020, Austria (email: \texttt{alessandro.crismani@gmail.com})}, Stavros Toumpis\thanks{Stavros Toumpis is with the Athens University of Economics and Business, Department of Informatics, Athens, Greece}
, Udo Schilcher, G\"{u}nther Brandner, and Christian Bettstetter\thanks{This work has been supported by the Austrian Science Fund (FWF) under grant P24480-N15. It has also been supported by the ERDF, KWF, and state of Austria under grants KWF-20214/15935/23108 (RELAY) and 20214/20777/31602 (Research Days) within the research cluster Lakeside Labs. The work of S. Toumpis was supported by the European Union (European Social Fund --- ESF) and Greek national funds through the Operational Program ``Education and Lifelong Learning'' of the National Strategic Reference Framework (NSRF) --- Research Funding Program: ``THALES DISCO - Investing in knowledge society through the European Social Fund.''}

}

\markboth{}{}
\thispagestyle{empty}

\newcounter{mytempeqncnt}

\newcommand{\etal}{\textit{et al.\;}}
\newcommand{\expect}{{\rm E}}
\newcommand{\prob}{{\rm P}}
\newcommand{\indic}{{\rm\bf 1}}
\newcommand{\dd}{\:\:{\rm d}}
\newcommand{\eul}{\mathrm{e}}
\newcommand{\ppp}{\Phi}
\newcommand{\pppt}{\Psi}
\newcommand{\txp}{p}
\newcommand{\recp}{S_d}
\newcommand{\dens}{\lambda}
\newcommand{\thr}{\theta}
\newcommand{\thri}{\theta_i}
\newcommand{\tn}{t}
\newcommand{\tlim}{{T}}
\newcommand{\edge}{L}
\newcommand{\src}{a}
\newcommand{\dst}{b}
\newcommand{\cluster}{\mathcal{R}}
\newcommand{\loss}{g}
\newcommand{\losssd}{g_{\src \dst}}
\newcommand{\lossud}{\loss_{u \dst}}
\newcommand{\chsd}{h_{\src \dst}}
\newcommand{\chud}{h_{u \dst}}
\newcommand{\channel}{h}
\newcommand{\rxd}{P_\dst}
\newcommand{\intd}{I_\dst}
\newcommand{\sir}{\mathrm{SIR}_{\src \dst}}

\newcommand{\source}{s}
\newcommand{\relay}[1]{r_{{#1}}}
\newcommand{\destination}{d}
\newcommand{\chSD}{h_{\source \destination}}
\newcommand{\pdfchSD}{\eul^{-\chSD}}
\newcommand{\chSR}[1]{h_{\source {#1}}}
\newcommand{\chRD}[1]{h_{{#1} \destination}}
\newcommand{\chuD}{h_{u \destination}}
\newcommand{\chuR}[1]{h_{u {#1}}}
\newcommand{\rxD}{P_{\source\destination}}
\newcommand{\intD}{I_\destination}
\newcommand{\intR}[1]{{I_{#1}}}
\newcommand{\ualoha}{\mathbf{1}_u}
\newcommand{\lossSD}{\loss_{\source \destination}}
\newcommand{\lossSR}[1]{\loss_{\source {#1}}}
\newcommand{\lossRD}[1]{\loss_{{#1} \destination}}
\newcommand{\lossuD}{\loss_{u \destination}}
\newcommand{\lossuR}[1]{\loss_{u {#1}}}
\newcommand{\lossxD}{\loss_{x \destination}}
\newcommand{\lossxR}[1]{\loss_{x {#1}}}
\newcommand{\thrSD}{{\thr_{\source \destination}}}
\newcommand{\thrSR}[1]{{\thr_{\source {#1}}}}
\newcommand{\thrRD}[1]{{\thr_{{#1} \destination}}}
\newcommand{\sirSD}{\rho_{\source \destination}}
\newcommand{\sirSR}[1]{\rho_{\source {#1}}}
\newcommand{\sirRD}[1]{\rho_{{#1} \destination}}
\newcommand{\sirSRD}[1]{\rho_{\source {#1} \destination}}
\newcommand{\coopgain}{\eta}
\newcommand{\chSDt}[1]{h_{\source \destination}^{#1}}
\newcommand{\intDt}[1]{\sum_{u \in \ppp} h_{u \destination}^{#1} \lossuD \ualoha}
\newcommand{\card}[1]{|{#1}|}
\newcommand{\powerset}[1]{\mathcal{P}\left( {#1} \right)}
\newcommand{\Esucc}{\mathcal{S}}
\newcommand{\Esd}{S_{0}}
\newcommand{\Esrd}[1]{S_{#1}^{\mathrm{SC}}}
\newcommand{\Esrdplain}[1]{S_{#1}}
\newcommand{\EsrdMRC}[1]{S_{#1}^{\mathrm{MRC}}}
\newcommand{\cevT}[1]{\overline{S^{#1}}}
\newcommand{\evT}[1]{S^{#1}}
\newcommand{\evSDt}[1]{S_{\source \rightarrow \destination}^{#1}}
\newcommand{\evSRt}[1]{S_{\source \rightarrow \relay{}}^{#1}}
\newcommand{\evRDt}[1]{S_{\relay{} \rightarrow \destination}^{#1}}
\newcommand{\cevSDt}[1]{\overline{S_{\source \rightarrow \destination}^{#1}}}
\newcommand{\cevSRt}[1]{\overline{S_{\source \rightarrow \relay{}}^{#1}}}
\newcommand{\cevRDt}[1]{\overline{S_{\relay{} \rightarrow \destination}^{#1}}}
\newcommand{\Pij}{\prob_{i,j}}
\newcommand{\Psucc}{\Omega}
\newcommand{\throughput}{\Lambda}
\newcommand{\PsExactT}[1]{\prob_S^{#1}}
\newcommand{\PsPhase}[1]{\prob_\mathrm{phase}\left({#1}\right)}
\newcommand{\PsPhaseAttempt}[2]{\prob_\mathrm{stage}\left({#1},{#2}\right)}
\newcommand{\Psd}{\prob\left[ \Esd \right]}
\newcommand{\Psrd}[1]{\prob\left[ \Esrd{#1} \right]}
\newcommand{\speed}{v}
\newcommand{\corr}{f_d D_0}
\newcommand{\gap}{B}
\newcommand{\efficiency}{\xi}
	
\sloppy

\maketitle

\thispagestyle{empty}

\begin{abstract}
We analyze the performance of an interference-limited, decode-and-forward, cooperative relaying system that comprises a source, a destination, and $N$ relays, placed arbitrarily on the plane and suffering from interference by a set of interferers placed according to a spatial Poisson process. In each transmission attempt, first the transmitter sends a packet; subsequently, a single one of the relays that received the packet correctly, if such a relay exists, retransmits it. We consider both selection combining and maximal ratio combining at the destination, Rayleigh fading, and interferer mobility. 

We derive expressions for the probability that a single transmission attempt is successful, as well as for the distribution of the transmission attempts until a packet is transmitted successfully. Results provide design guidelines applicable to a wide range of systems. Overall, the temporal and spatial characteristics of the interference play a significant role in shaping the system performance. Maximal ratio combining is only helpful when relays are close to the destination; in harsh environments, having many relays is especially helpful, and relay placement is critical; the performance improves when interferer mobility increases; and a tradeoff exists between energy efficiency and throughput.

\end{abstract}

\section{Introduction}
\label{sec:intro-related}
\PARstart{T}{he} properties of interference have a significant impact on the performance of wireless systems~\cite{haenggi09:interference,4927481}. In broad terms, a transmitter-receiver pair suffers from interference if one or more 
signals from other transmitters add up with the useful signal at the receiver, causing decoding errors, necessitating the use of lower data rates, and ultimately leading to a reduction of the overall network capacity~\cite{Hekmat:2004}.

The sum of the powers of all interfering signals at the location of the receiver, which we call the interference power at that location, critically affects the decoding; its expected value is an important parameter that must be properly considered when designing a wireless system. However, the manner in which interference power changes across time and space should also be considered, especially in the design of temporal and spatial diversity schemes~\cite{haenggi09:outage,ganti09:interf-correl,net:Zhong13twc,tanbourgi2013coop}. We refer to the laws that govern the fluctuation of the interference power across time and space as the \textbf{interference dynamics}.

This article applies methods from the theory of spatial stochastic processes to create accurate interference dynamics and study the effect of these on cooperative relaying (also called cooperative diversity)~\cite{Laneman04cooperativediversity}. Most research so far analyzed cooperative relaying without a careful consideration of interference dynamics, and our work bridges this gap.

We consider a decode-and-forward cooperative relaying system comprised of a source, a destination, and $N$ relays. The source sends packets to the destination using consecutive two-slotted transmission attempts: in the first slot of each attempt, the source transmits the packet. In the second slot, one of the relays that received the packet, if such a relay exists, retransmits it. The destination attempts to decode the signal in both time slots; in the second time slot it uses either selection combining or maximal ratio combining of the two signals received from the source (in the first slot) and the relay (in the second slot). If the transmission attempt fails, another one is initiated at a later time. All links are subject to path loss and Rayleigh fading.

Observe that the cooperative relaying scheme we employ is straightforward; the major novelty of this work lies in the analysis of the system under interference dynamics. In particular, we assume that interferers are distributed according to a spatial Poisson process. Three mobility models are considered for them, according to which their locations at different transmission attempts are fixed (modeling stationary interferers), independent (modeling highly mobile interferers), or correlated (modeling interferers with intermediate levels of mobility). 

In this setting, we find the probability that a single transmission attempt is successful as well as the distribution of the number of transmission attempts until a packet is successfully transmitted. Our work offers several insights, some expected, some not so expected:
\begin{itemize}
\item Cooperative relaying, even with a single available relay, can significantly improve the performance of the system. However, employing numerous relays only makes sense in particularly harsh environments with increased interference and lack of reliable links between the source and the destination.
\item The best locations for placing the relays are in-between the source and the destination, with a small bias towards the destination. Their placement affects the performance more drastically in harsh scenarios.
\item Maximal ratio combining is only beneficial when the relays are placed much closer to the source than to the destination.
\item The performance improves as interferers become more mobile and, hence, the success events at consecutive attempts more uncorrelated.
\item When interferers are mobile, we improve the energy efficiency of the system by reducing its throughput, and vice versa.
\end{itemize}

Overall, the contributions of this work improve our understanding of the performance of wireless systems in the presence of interference, providing insights on how interference dynamics affects the packet delivery probability and the system throughput. Hence, they are a step forward in the design of wireless transmission schemes that operate efficiently in interference-limited environments.

The rest of this paper is organized as follows. Section~\ref{sec:related} reviews the related literature and provides background for our study. Section~\ref{sec:modelling} describes the system under study and details the modeling assumptions. The packet delivery probability for a single transmission attempt is derived in Section~\ref{sec:pkt-delivery}. An ARQ scheme, where multiple transmission attempts take place until there is a successful one, is analyzed in Section~\ref{sec:arq}. Finally, Section~\ref{sec:conclusions} summarizes the main results and outlines possible future extensions.

We note that a preliminary version of part of this work (notably, the results shown in Section~\ref{sec:onerelay} and a part of the results provided in Section~\ref{sec:arq}) appeared in~\cite{schilcher2013coop}. 

\section{Related Work}
\label{sec:related}

In cooperative relaying schemes, the communication between a source-destination pair is supported by one or more cooperating relays. In recent years, starting with the seminal work in~\cite{Laneman04cooperativediversity}, where the authors designed and compared a wide range of practical relaying schemes and showed that cooperation provides a significant resource gain compared to non-cooperative solutions, the performance of cooperative relaying has been studied intensely.

Despite the extend of research efforts on cooperative relaying, only few papers focus on the effect of interference on the performance of such schemes. Notably, the capacity of a communication system where two transmitter-receiver pairs operate simultaneously, mutually causing interference, is analyzed in~\cite{peng2009twouser}. There, cooperation is enabled by letting the two transmitters help each other and as a result the proposed cooperative scheme increases the network performance. The tradeoff between the benefit of adopting cooperative relaying and the interference generated by relays is highlighted in~\cite{liang2008coopinterf} in terms of the sum-rate and energy efficiency of a cooperative asynchronous multi-user scenario. Decode-and-forward relaying schemes where cooperative transmissions incur interference are studied in~\cite{ikki2013coopinterf}. There, the authors derive the outage probability and provide optimal energy allocation strategies for an adaptive relay selection scheme. 

Although these investigations take into account the influence of interfering nodes on cooperative communications, they do not use accurate models for the interference dynamics. However, as we show in this work, the interference model adopted significantly affects the derived performance.

A number of recent publications have focused on the effects of interference dynamics on the performance of wireless networks, specifically in the case where the interferers are placed on the plane according to a spatial Poisson process. In particular, Haenggi derives the outage probability, probabilistic throughput, and ergodic capacity of network scenarios represented by the vertices of an `uncertainty cube'~\cite{haenggi09:outage}, which models the main sources of uncertainty and correlation in the network. A wider range of correlation sources is considered by some of the authors of the article at hand in~\cite{schilcher12:interfcor}, where the results of \cite{haenggi09:outage} are extended by considering various models for the nodes' locations, the temporal properties of fading, and the traffic pattern of the nodes. Similar interference dynamics are studied in~\cite{ganti09:interf-correl} for obtaining the conditional probability of outage in a network where different transmissions are affected by the same set of interferers. The analysis is extended to the cooperative domain in~\cite{schilcher2013coop}, where the authors of the article at hand analyze the performance of a single-hop cooperative system with one relay and a two-hop cooperative system with two relays under the influence of interference powers that are correlated across time and space. A similar system is studied in~\cite{tanbourgi2013coop}, where the authors derive the outage probability of a relaying scheme where the destination combines the signal obtained from the source and from either a second source transmission or a relay transmission;~\cite{tanbourgi2013coop} also discusses the diversity order of that cooperative system. The diversity order is also studied in \cite{haenggi12ltt}, in a Single-Input Multiple-Output (SIMO) setting. More recently, the authors of \cite{tanbourgi14a}, \cite{tanbourgi14b} study the effects of interference dynamics on the performance of maximal ratio combining.

The work at hand follows this line of research, i.e., studies cooperative relaying where the interference dynamics comes from modeling the placement of interferers according to a spatial Poisson process. However, we consider multiple relays and multiple packet retransmissions. Furthermore, we also study the effect of mobility by providing analytical results for two mobility models specifying the placement of interferers across different transmission attempts and simulation results for a third mobility model that bridges the gap between the other two. These contributions represent an advance towards a better theoretical understanding of cooperative relaying systems, and help in the efficient design of practical cooperative relaying systems operating in interference-limited scenarios.

\section{Modeling Assumptions}
\label{sec:modelling}
\subsection{Communication Scheme}
\label{sub:communication_scheme}
We study a decode-and-forward cooperative relaying system where the communication between a source $\source$ and a destination $\destination$ is aided by a set of $N$ relays $\left\{ \relay{n} \right\}_{n=1}^N$. Let $\source, \destination$, and $\relay{n}$, $n=1,\ldots,N$, denote both these nodes and their locations on the plane, which do not change with time.

We adopt the following time-slotted cooperative relaying scheme: the source transmits a packet during a particular time slot; all relays and the destination attempt to decode the transmission. Following the commonly adopted approach of, e.g.,~\cite{liu2007coopmac}, where a period of time right after the source's transmission is reserved for relays' transmissions, we assume that one of the relays that decoded the packet correctly, if one exists, forwards the packet to the destination in the immediately following time slot. We refer to the operation of the nodes during these two time slots as a \textbf{transmission attempt} (or simply \textbf{attempt}).

If the destination does not correctly decode the packet in one of the two time slots, a new attempt, following the same strategy, is carried out, so that the time that passes between the starts of consecutive attempts is equal to the \textbf{inter-attempt time} $D_0$. Intuitively, $D_0$ models the time needed for the source to regain access to the channel through a medium access protocol, and might be much larger than the duration of a single slot. 

Therefore, we have constrained our work to the case where only a single node transmits the packet at any time slot. This transmission strategy is not optimal, since allowing multiple nodes to simultaneously transmit might lead to a higher packet delivery probability. However, a system where multiple nodes simultaneously transmit requires tight node synchronization and possibly also the adoption of more advanced coding techniques, such as distributed space-time block coding. Furthermore, it was shown in~\cite{bletsas2006simple} that a carefully constructed system where a single, properly chosen relay forwards the packet received from the source provides the same diversity order as the one attained by a system where multiple relays transmit simultaneously.

Also, we follow~\cite{bletsas2006simple} and assume that if there is at least one relay that can support the communication, the source-destination pair can identify it and invoke its help. For the case where there are multiple such relays, various methods for selecting a particular one have been proposed; they include, notably, contention between relays~\cite{shan2009mac, adam2014} and selection based on a table that stores information about the quality of candidate relays~\cite{liu2007coopmac,zhu2006rdcf}. We dispense with specifying a particular method, since the analysis of the relay selection phase is outside the scope of this work.

\subsection{Interference and Interferer Mobility Models}
\label{sub:locations_of_interferers}

Transmissions may not be received successfully because they are subject to co-channel interference. In particular, we assume a set of (possibly mobile) interferers distributed on the plane, at any given time, according to a Poisson point process (PPP) $\ppp$ of intensity $\dens$~\cite{stoyan95}. Let $u \in \ppp$ denote both a generic interferer as well as its location.

We assume that each interferer transmits with probability $p$ during each time slot independently of the other interferers. We further assume that interferers that are active during the source's transmission are also active during the relay's transmission in the subsequent time slot. This assumption is motivated by the fact that, in most cases of interest, interfering nodes are oblivious to the communication scheme they interfere with.

Furthermore, we assume that in each transmission attempt each interferer will be transmitting independently of whether or not it transmitted in the previous ones. This is because consecutive transmission attempts are separated by multiple time slots, therefore the memory of whether a particular interferer has transmitted during an attempt is lost by the time the next attempt starts.

Regarding the movement of interferers, we study analytically two mobility models. The first model, which we call the Stationary Interferer Model (SIM), assumes that the interferer locations do not change over time, i.e., they remain fixed at all time slots and transmission attempts, thus following a single realization of the PPP $\ppp$. Clearly, this model is appropriate in the case of wireless networks where the interferers do not move, or move very slowly.

The second model, which we call the Meteoric Interferer Model (MIM), assumes that the positions of interferers during a transmission attempt remain fixed, but follow independent realizations of $\ppp$ during different transmission attempts. This model captures the scenario in which the interferers are highly mobile so that their locations during one transmission attempt do not provide information about their locations during any other transmission attempt, as these attempts are separated by the inter-attempt time $D_0$. On the other hand, as each transmission attempt comprises only two time slots, it is reasonable to assume that the locations of interferers remain constant during these.

Clearly, these two models represent two opposite extremes in the mobility of interferers. For this reason, apart from providing analytical results for them, we also provide simulation results for a third mobility model that bridges the gap between them. 

The third mobility model, which we refer to as the Traveling Interferer Model (TIM), is defined as follows: Interferer locations remain fixed during transmission attempts, but change from one attempt to the next. This change in the location of each interferer is determined by sampling an underlying continuous mobility model under which interferers move along straight lines with a constant speed $\speed$, common for all, and a direction of travel randomly and independently chosen for each interferer, uniformly in the interval $[0, 2 \pi)$. Therefore, an interferer moves a total distance of $vD_0 $ from one transmission attempt to the next. Results for this model are obtained only through Monte Carlo simulations.

\subsection{Channel Model}
\label{sub:channel_model}
\begin{figure}
    \centering
    \begin{tikzpicture}[auto]
        \pgfplotsset{tick style={draw=none}}
            \node[black, draw, circle, minimum size=0.7cm, inner sep=1pt] at  (0,0) (source) {$\source$};
            \node[black, draw, circle, minimum size=0.7cm, inner sep=1pt] at (6,0) (destination) {$\destination$};
            \node[draw, circle, minimum size=0.7cm, inner sep=1pt] at (2,2) (r1) {$\relay{1}$};
            \node[draw, circle, minimum size=0.7cm, inner sep=1pt] at (3.5,-2.3) (r2) {$\relay{2}$};
           \draw[->, thick] (source) -- node[midway] {$\chSD \; \lossSD$} (destination);
           \draw[->, thick, dashed] (source) -- node[near end, sloped] {$\chSR{1} \; \lossSR{1}$} (r1);
           \draw[->, thick, dashed] (r1) -- node[near start, sloped] {$\chRD{1} \; \lossRD{1}$} (destination);
           \draw[->, thick, dotted] (source) -- node[near start, sloped] {$\chSR{2} \; \lossSR{2}$} (r2);
           \draw[->, thick, dotted] (r2) -- node[near end, sloped] {$\chRD{2} \; \lossRD{2}$} (destination);
    \end{tikzpicture}
    \caption{Example network scenario for $N = 2$ relays.}
    \label{fig:scenario}
\end{figure}
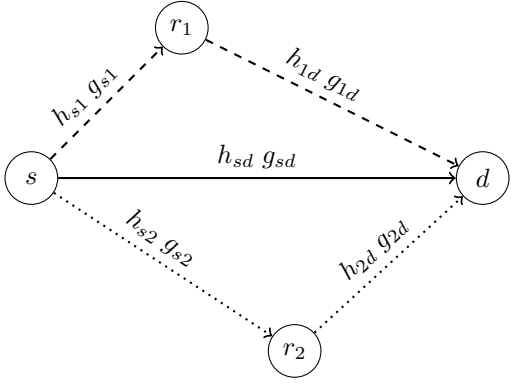

The channel is modeled assuming path loss combined with Rayleigh fading. In particular, the power received at $\destination$ when $\source$ transmits with power $P_s$ is
\begin{equation}
    \rxD = \chSD \lossSD P_s,
    \label{eqn:rx_power}
\end{equation}
where the fading coefficient $\chSD$ models Rayleigh fading and is an exponentially distributed random variable with mean equal to unity\footnote{Note that in the case of Rayleigh fading, the \emph{amplitude} of the signal is Rayleigh distributed, and therefore the fading coefficient appearing in~\eqref{eqn:rx_power}, which is proportional to the received \emph{power}, is exponentially distributed.}, and the strictly positive path loss coefficient $\lossSD$ represents path loss. Without loss of generality, we assume that all transmitter powers are equal to unity. Similarly, $\chSR{n}$, $\chRD{n}$, $\lossSR{n}$ and $\lossRD{n}$ denote the fading and path loss coefficients of the links connecting $\source$ to $\relay{n}$ and $\relay{n}$ to $\destination$. We also denote the fading and path loss coefficients of the links connecting $u$ to $\relay{n}$ and $u$ to $\destination$ with $\chuR{n}$, $\chuD$, $\lossuR{n}$ and $\lossuD$. Fig.~\ref{fig:scenario} shows an example network with $N = 2$ relays.

All fading coefficients remain constant for the duration of a transmission attempt, i.e., the two consecutive time slots hosting the transmissions of the source and the relay. We make this assumption since we have to assume, for reasons of mathematical tractability, that fading coefficients are fixed for the full duration of a single time slot; therefore, it is reasonable to assume that they will also not change in the immediately following slot. 
We note that analytical results  when the fading coefficients over two consecutive time slots are independent are reported, for a related setting, in~\cite{tanbourgi2013coop}.

We also assume that the fading coefficients of the same link at different transmission attempts are independent. Also, the fading coefficients of different links at the same or at different transmission attempts are independent, even if these links share a single common node.

We do not adopt any particular model for the path loss coefficients, and the expressions presented in this work are valid for any model. In deriving all our numerical results, we will use the following path loss model:
\begin{equation}
    \lossSD = \| \source - \destination \| ^ {-\alpha},
    \label{eqn:path_loss}
\end{equation}
where the path loss exponent $\alpha$ is set to $\alpha = 4$. (Similar expressions hold for all other transmitter-receiver pairs.)
Note that --- as we assumed that the source, relays and the destination do not move --- the path loss coefficients $\lossSD$, $\lossSR{n}$ and $\lossRD{n}$ do not vary with time. On the other hand, depending on the mobility model, the values of the path coefficients $\lossuR{n}$ and $\lossuD$ might vary across transmission attempts.

\subsection{Receiver Model}
\label{sub:receiver_model}

We assume that communication is interference-limited, and hence we neglect the effects of noise. The analysis can be easily extended to include them, when the fading amplitudes follow the Rayleigh distribution~\cite{haenggi09:interference}.

The transmission technology adopted is such that the relays and destination correctly decode a transmitted packet if and only if the signal to interference ratio (SIR) at their receiver is higher than a threshold $\thr$.

Finally, we consider two decoding rules at the destination, namely selection combining (SC) and maximal ratio combining (MRC). Under both of them the destination attempts to decode the transmission of the source in the first time slot and the transmission of the relay (if a relay transmits) in the second time slot. However, under MRC, in the second time slot the destination adds the power received from the source in the first time slot to the power received from the relay, while under SC it uses solely the power received from the relay.

Having specified the receiver model, we proceed to define quantities and events related to a single transmission attempt that will be used in the subsequent analysis.

First, let the indicator function
\begin{equation}
\ualoha=
\begin{dcases*}
1, & if interferer $u$ transmits,\\
0, & otherwise.
\end{dcases*}
\end{equation}
Therefore, the interference power at $\destination$, in both time slots of the transmission attempt, equals
\begin{equation}
    \intD = \sum_{u \in \ppp} \chuD \lossuD \ualoha.
    \label{eqn:interference}
\end{equation}
Similarly, the interference power at the $n$-th relay, in the first time slot, equals
\begin{equation}
    \intR{n} = \sum_{u \in \ppp} \chuR{n} \lossuR{n} \ualoha.
    \label{eqn:interference_relay}
\end{equation}

In the case of SC, the SIR $\sirSD$ at the destination $\destination$ of the signal transmitted by the source $\source$ during the first time slot is
\begin{equation}
    \sirSD = \frac{\chSD \lossSD}{\intD}=\frac{\chSD \lossSD}{\sum_{u \in \ppp} \chuD \lossuD \ualoha}.
    \label{eqn:sir}
\end{equation}
Similar expressions, \emph{mutatis mutandis}, characterize the SIR $\sirSR{n}$ at the relay $\relay{n}$ of the signal transmitted by the source $\source$ during the first time slot and the SIR $\sirRD{n}$ at the destination $\destination$ of the signal transmitted by the relay $\relay{n}$ during the second time slot (provided that the relay transmits).

\begin{figure*}[!t]
    \setcounter{mytempeqncnt}{\value{equation}}
    \setcounter{equation}{10}
    \begin{align}
        \prob[A] &= \prob[\Esd \cap \Esrd{1} \cap \dots \cap \Esrd{K}] \notag \displaybreak[0] \\
        &= \prob\left[ \chSD \lossSD > \thr \intD, \chSR{1} \lossSR{1} > \thr \intR{1}, \chRD{1} \lossRD{1} > \thr \intD, \dots, \chSR{K} \lossSR{K} > \thr \intR{K}, \chRD{K} \lossRD{K} > \thr \intD \right] \notag \displaybreak[0] \\
        &\overset{(a)}{=} \expect_{\ppp, \channel, \ualoha} \left[ \eul^ {\left( - \thrSD \sum \limits_{u \in \ppp} \chuD \lossuD \ualoha \right)} \prod_{k=1}^K \left ( \eul ^ {\left( - \thrSR{k} \sum \limits_{u \in \ppp} \chuR{k} \lossuR{k} \ualoha \right)} \eul ^ {\left( - \thrRD{k} \sum \limits_{u \in \ppp} \chuD \lossuD \ualoha \right)} \right) \right] \notag \displaybreak[0] \\
        &= \expect_{\ppp, \channel, \ualoha} \left[ \prod_{u \in \ppp} \left ( \eul ^ {\left( - \thrSD \chuD \lossuD \ualoha \right)} \prod_{k=1}^K \left( \eul ^ {\left( - \thrSR{k} \chuR{k} \lossuR{k} \ualoha \right)} \eul ^ {\left( - \thrRD{k} \chuD \lossuD \ualoha \right)} \right) \right) \right] \notag \displaybreak[0] \\
        &\overset{(b)}{=} \expect_{\ppp} \left[ \prod_{u \in \ppp} \expect_{\ualoha} \left[ \expect_{\chuD} \left[ \eul ^ {\left( - \left( \thrSD + \sum \limits_{k=1}^K \thrRD{k} \right) \chuD \lossuD \ualoha \right)} \right] \prod_{k=1}^K \expect_{\chuR{k}} \left[ \eul ^ {- \left( \thrSR{k} \chuR{k} \lossuR{k} \ualoha \right)} \right] \right] \right] \notag \displaybreak[0] \\
        &\overset{(c)}{=} \expect_{\ppp} \left[ \prod_{u \in \ppp} \left( \left( \frac{\txp}{1 + \left( \thrSD + \sum \limits_{k=1}^K \thrRD{k} \right) \lossuD} \prod_{k=1}^K \frac{1}{1 + \thrSR{k} \lossuR{k}} \right) + 1 - \txp \right) \right] \notag \displaybreak[0] \\
        &\overset{(d)}{=} \exp \left( -\dens \bigints\limits_{\mathbb{R}^2} \left[ 1 - \left( \left( \frac{\txp}{1 + \left( \thrSD + \sum \limits_{k=1}^K \thrRD{k} \right) \lossxD} \prod_{k=1}^K \frac{1}{1 + \thrSR{k} \lossxR{k}} \right) + 1 - p \right) \right] \dd x \right).
        \label{eqn:dir_link_sel}
    \end{align}
    \setcounter{equation}{\value{mytempeqncnt}}
    \hrule
\end{figure*}

In the case of MRC, all SIRs are given by the same expressions as in the SC case, except for the SIR at the destination  $\destination$ during the second time slot in case the $n$-th relay was selected to support the communication. In this case, the SIR equals
\begin{equation}
    \sirSRD{n} = \frac{\chSD \lossSD + \chRD{n} \lossRD{n}}{\intD} = \frac{\chSD \lossSD + \chRD{n} \lossRD{n}}{\sum_{u \in \ppp} \chuD \lossuD \ualoha}.
    \label{eqn:sir_mrc}
\end{equation}

Next, let $\Esd$ be the event of successful decoding at $\destination$ of the signal transmitted by $\source$, for both SC and MRC. We have
\begin{equation}
    \Esd = \{ \sirSD > \thr\}.
    \label{eqn:success_sd}
\end{equation}
Similarly, let $\Esrd{n}$ be the event of a successful packet delivery using $\relay{n}$ and SC at $\destination$. This event corresponds to a successful decoding of the signal at $\relay{n}$, followed by a successful decoding at $\destination$ of the signal received from $\relay{n}$. Hence,
\begin{equation}
    \Esrd{n} =\{ \sirSR{n} > \thr \wedge \sirRD{n} > \thr\}.
    \label{eqn:success_sel}
\end{equation}
A similar definition can be used when $\destination$ adopts MRC. In particular, let $\EsrdMRC{n}$ be the event that $\relay{n}$ correctly decodes the signal transmitted by $\source$, and $\destination$ successfully decodes the combination of the signals received from $\source$ and $\relay{n}$. We have
\begin{equation}
    \EsrdMRC{n} =\{ \sirSR{n} > \thr \wedge \sirSRD{n} > \thr\}.
    \label{eqn:success_mrc}
\end{equation}

Finally, we denote the ratio between the success threshold $\thr$ and the path loss coefficient between $\source$ and $\destination$ by
\addtocounter{equation}{1}
\begin{equation}
    \thrSD = \frac{\thr}{\lossSD}.
\end{equation}
Similar definitions apply to all other transmitter-receiver pairs.

\section{Single Transmission Attempt}
\label{sec:pkt-delivery}
\subsection{Selection Combining}
\label{sec:sel}
We define the \textbf{success probability} (\textbf{SP}) $\Psucc$ of a single transmission attempt to be the probability that the direct transmission (in the first time slot) is successful or both the transmissions of one of the relay-aided $2$-hop paths (occupying the two consecutive time slots) are successful. Hence,
\begin{equation}
    \Psucc = \prob \left[ \Esd \cup \left( \bigcup_{n=1}^N \Esrd{n} \right) \right].
    \label{eqn:succ_union}
\end{equation}
The outage probability, i.e. the probability that the packet is not received correctly due to an unfavorable combination of poor fading conditions and interference, is equal to one minus the success probability. For simplicity, we will not use outage probability in the following discussions.

By applying the inclusion-exclusion principle we get
\begin{equation}
    \Psucc = \sum_{A \in \powerset{\Esucc}} (-1)^{\card{A} + 1} \prob[A],
    \label{eqn:succ_incl_excl}
\end{equation}
where $\Esucc = \{ \Esd, \Esrd{1}, \dots, 
\Esrd{N} \}$, $\powerset{\Esucc}$ is the power set of $\Esucc$ excluding the empty set, $\card{A}$ denotes the cardinality of the set $A$, and, finally, $\prob[A]$ is the probability of the intersection of all events in set $A$. It follows from~\eqref{eqn:succ_incl_excl} that, in order to find $\Psucc$, it suffices to find the probabilities $\prob [A],~ \forall A \in \powerset{\Esucc}$. These probabilities are derived in the following. The analysis is divided into two cases: the case where set $A$ contains $\Esd$, and the case where $A$ does not contain $\Esd$.
\begin{figure*}[!t]
    \setcounter{mytempeqncnt}{\value{equation}}
    \setcounter{equation}{14}
    \begin{align}
        &\prob[A] \overset{(a)}{=} \prob[\EsrdMRC{1} \cap \dots \cap \EsrdMRC{K}] = \prob \left[ \chSR{1} \lossSR{1} > \thr \intR{1}, \chSD \lossSD + \chRD{1} \lossRD{1} > \thr \intD, \dots, \chSR{K} \lossSR{K} > \thr \intR{K}, \chSD \lossSD + \chRD{K} \lossRD{K} > \thr \intD \right] \notag \displaybreak[0] \\
        &\overset{(b)}{=} \coopgain \expect_{\ppp, \channel, \ualoha} \left[\prod_{k=1}^K \mathrm{e} ^ { - \thrSR{k} \sum \limits_{u \in \ppp} \chuR{k} \lossuR{k} \ualoha } \eul ^ { -\thrRD{k} \sum \limits_{u \in \ppp} \chuD \lossuD \ualoha } \right] + \left(1 \! - \! \coopgain \right) \expect_{\ppp, \channel, \ualoha} \left[\prod_{k=1}^K \eul ^ { \left( - \thrSR{k} \sum \limits_{u \in \ppp} \chuR{k} \lossuR{k} \ualoha \right) }  \eul ^ { \left( -\thrSD \sum \limits_{u \in \ppp} \chuD \lossuD \ualoha \right) } \right] \notag \displaybreak[0] \\
        &=  \coopgain \expect_{\ppp} \!\! \left[\prod_{u \in \ppp} \left( \!\! \left( \frac{\txp}{1 + \sum \limits_{k=1}^K \thrRD{k} \lossuD} \prod_{k=1}^K \frac{1}{1 + \thrSR{k} \lossuR{k}} \right) + 1 - \txp \right) \! \right] +  \left(1 - \coopgain \right) \expect_{\ppp} \!\! \left[\prod_{u \in \ppp} \left( \!\! \left( \frac{\txp}{1 + \thrSD \lossuD} \prod_{k=1}^K \frac{1}{1 + \thrSR{k} \lossuR{k}} \right) + 1 - \txp \right) \! \right] \notag \displaybreak[0] \\
        &=  \coopgain \exp \! \left( \! - \! \dens \! \bigintss_{\mathbb{R}^2} \! \left(1 - \left( \left( \frac{\txp}{1 + \sum \limits_{k=1}^K \thrRD{k} \lossxD} \prod_{k=1}^K \frac{1}{1 + \thrSR{k} \lossxR{k}} \right) + 1 - \txp  \right) \right) \dd x \right) + \notag \\ &\quad \left(1 - \coopgain \right) \exp \left( -\dens \bigintsss_{\mathbb{R}^2} \left(1 - \left( \left( \frac{\txp}{1 + \thrSD \lossxD} \prod_{k=1}^K \frac{1}{1 + \thrSR{k} \lossxR{k}} \right) + 1 - \txp  \right)\right) \dd x \right).
        \label{eqn:coop_link_mrc}
    \end{align}
    \setcounter{equation}{\value{mytempeqncnt}}
    \hrule
\end{figure*}

Firstly, consider the set of events $A = \{ \Esd, \Esrd{1}, \dots, \Esrd{K}\}$ comprising the events of success on the source-destination link and on the first $K$ $2$-hop paths, where $0 \leq K \leq N$. Note that the following analysis holds for any set composed by $\Esd$ and any $K$ events corresponding to the successful use of $K$ $2$-hop paths. However, to keep the notation simple, the result is presented in~\eqref{eqn:dir_link_sel} for the first $K$ $2$-hop paths.

In~\eqref{eqn:dir_link_sel}, in $(a)$ we condition on the realization of $\ppp$, the fading coefficients of the links involving the interferers (and \emph{only} on them), and whether the interferers are active or not. The only sources of randomness left are the fading coefficients $\chSD, \left\{ \chSR{k} \right\}_{k=1}^K, \left\{ \chRD{k} \right\}_{k=1}^K$, which follow the exponential distribution with unit mean and are independent of each other, hence the resulting expression; $(b)$ follows from the fact that the fading coefficients of different links are independent, and because each interferer decides to transmit independently of the others; $(c)$ follows from first calculating the expectations over the fading coefficients through the use of the characteristic function of the exponential distribution and then taking the expectations over the indicator functions $\ualoha$; $(d)$ is obtained by applying the probability generating functional of $\ppp$~(cf.~\cite{haenggi13:book}, (4.8)).

Secondly, consider the set $A = \{ \Esrd{1}, \dots, \Esrd{K}\}$. We have
\addtocounter{equation}{1}
\begin{align}
   \prob[A] &= \prob \left[\Esrd{1} \cap \dots \cap \Esrd{n} \right] \notag \\ &= \prob\Big[ \chSR{1} \lossSR{1} > \thr \intR{1}, \chRD{1} \lossRD{1} > \thr \intD, \dots, \notag \\ &\chSR{K} \lossSR{K} > \thr \intR{K}, \chRD{K} \lossRD{K} > \thr \intD \Big] \notag \\
    &\overset{(a)}{=} \exp \left(-\dens \int_{\mathbb{R}^2} \left[ 1 - \left( \left( \frac{\txp}{1 + \left(\sum_{k=1}^K \thrRD{k} \right) \lossxD} \cdot \notag \right. \right. \right. \right. \\ & \left. \left. \left. \left. \prod_{k=1}^K \frac{1}{1 + \thrSR{k} \lossxR{k}} \right) + 1 - p \right) \right] \dd x \right),
    \label{eqn:coop_link_sel}
\end{align}
where $(a)$ follows from the exact same steps as in~\eqref{eqn:dir_link_sel} without accounting for the contribution of the source-to-destination link.

Therefore, we can calculate $\Psucc$ using (\ref{eqn:succ_incl_excl}), calculating all terms appearing in (\ref{eqn:succ_incl_excl}) using (\ref{eqn:dir_link_sel}) and (\ref{eqn:coop_link_sel}).

\subsection{Maximal Ratio Combining}
\label{sec:mrc}
Next, we consider a scenario where the destination adopts MRC for jointly decoding the signals received from the source and from the selected relay. The SP $\Psucc$ now becomes
\begin{equation}
    \Psucc = \prob \left[ \Esd \cup \left( \bigcup_{n=1}^N \EsrdMRC{n} \right) \right],
    \label{eqn:succ_union_mrc}
\end{equation}
and hence one can use~\eqref{eqn:succ_incl_excl}, where now the probabilities $\prob[A]$ are calculated assuming MRC.

Firstly, let $A = \{ \Esd, \EsrdMRC{1}, \dots, \EsrdMRC{K} \}$. We have
\begin{align}
    &\prob[A] = \prob \left[ \Esd \cap \EsrdMRC{1} \cap \dots \cap \EsrdMRC{K} \right] \notag \\
    &= \prob\left[ \chSD \lossSD \! > \! \thr \intD, \chSR{1} \lossSR{1} \! > \! \thr \intR{1}, \chSD \lossSD + \chRD{1} \lossRD{1} \! > \! \thr \intD, \dots, \right. \notag \\ &\qquad \left. \chSR{K} \lossSR{K} \! > \thr \intR{K}, \chSD \lossSD + \chRD{K} \lossRD{K} \! > \! \thr \intD \right] \notag \\
    &\overset{(a)}{=} \prob\left[ \chSD \lossSD > \thr \intD, \chSR{1} \lossSR{1} > \thr \intR{1}, \dots, \chSR{K} \lossSR{K} > \thr \intR{K} \right] \notag \\
    &\overset{(b)}{=} \exp \left( -\dens \int_{\mathbb{R}^2} \left[ 1 - \left( \left( \frac{\txp}{1 + \thrSD \lossxD} \prod_{k=1}^K \frac{1}{1 + \thrSR{k} \lossxR{k}} \right) + \right. \right. \right. \notag \\ & \left. \left. \left. + 1 - p \right) \right] \dd x \right),
    \label{eqn:dir_link_mrc}
\end{align}
where $(a)$ follows from the fact that the condition $\chSD \lossSD > \thr \intD$ immediately implies the condition $\chSD \lossSD + \chRD{k} \lossRD{k} > \thr \intD$, $\forall k \in {1, \dots, K}$, and $(b)$ follows from applying steps similar to those of~\eqref{eqn:dir_link_sel}.
\begin{figure*}
    \setcounter{mytempeqncnt}{\value{equation}}
    \setcounter{equation}{19}
    \begin{align}
        \prob[\EsrdMRC{1}] &= \prob\left[ \chSR{1} \lossSR{1} > \thr \intR{1}, \chSD \lossSD + \chRD{1} \lossRD{1} > \thr \intD \right] \notag \\
        &=  \frac{1}{1 - \frac{\lossSD}{\lossRD{1}}} \exp \left( - \dens \int_{\mathbb{R}^2} \left[ 1 - \left( \frac{\txp}{\left( 1 + \thrRD{1} \lossxD \right) \left( 1 + \thrSR{1} \lossxR{1} \right)} +1 - \txp \right) \right] \dd x \right) + \notag \\ &\quad \left(1 - \frac{1}{1 - \frac{\lossSD}{\lossRD{1}}} \right) \exp \left( -\dens \int_{\mathbb{R}^2} \left[ 1 - \left( \frac{\txp}{\left( 1 + \thrSD \lossxD \right) \left( 1 + \thrSR{1} \lossxR{1} \right)} + 1 - \txp \right) \right] \dd x \right),
        \label{eqn:relay_link_onerelay_mrc}
    \end{align}
    \begin{align}
        \prob[\Esd \cap \EsrdMRC{1}] &= \prob\left[ \chSD \lossSD > \! \thr \intD, \chSR{1} \lossSR{1} > \! \thr \intR{1}, \chSD \lossSD + \chRD{1} \lossRD{1} > \! \thr \intD \right] \notag \\ &= \prob \left[ \chSD \lossSD > \thr \intD, \chSR{1} \lossSR{1} > \thr \intR{1} \right] \notag \\
        &= \exp \left( -\dens \int_{\mathbb{R}^2} \left[ 1 - \left( \frac{\txp}{\left( 1 + \thrSD \lossxD \right) \left( 1 + \thrSR{1} \lossxR{1} \right)} + 1 - p \right) \right] \dd x \right).
       \label{eqn:coop_link_onerelay_mrc}
    \end{align}
    \setcounter{equation}{\value{mytempeqncnt}}
    \hrule
\end{figure*}

Secondly, in order to obtain $P[A]$ for $A = \{ \EsrdMRC{1}, \dots, \EsrdMRC{K} \}$, consider at first the probability that the sum of the powers received at $\destination$ from $\source$ and from a single selected relay $\relay{k}$ is greater than a particular constant value $\beta$. This probability may be calculated as
\begin{align}
    &\prob \left[ \chSD \lossSD + \chRD{k} \lossRD{k} > \beta \right] = \expect_{\chSD} \left[ \prob \! \! \left[ \chRD{k} \! > \!\frac{\beta  - \chSD \lossSD}{\lossRD{k}} \right]  \right] \notag \\
    &= \int_{0}^{\frac{\beta}{\lossSD}} \eul ^ {\left( - \frac{\beta - \chSD \lossSD}{\lossRD{k}} \right)} \pdfchSD \dd \chSD + \int_{\frac{\beta}{\lossSD}}^{\infty} \pdfchSD \dd \chSD \notag \\
    &= \eul ^ {\left(-\frac{\beta}{\lossRD{k}} \right)} \int_{0}^{\frac{\beta}{\lossSD}} \eul ^ {\left( - \chSD \left(1 - \frac{\lossSD}{\lossRD{k}} \right) \right)} \dd \chSD + \eul ^ {\left( -\frac{\beta}{\lossSD} \right)} \notag \\
    &= \left( \frac{1}{1 - \frac{\lossSD}{\lossRD{k}}} \right) \eul ^ {\left(-\frac{\beta}{\lossRD{k}} \right)} + \left(1 - \frac{1}{1 - \frac{\lossSD}{\lossRD{k}}} \right) \eul ^ {\left( -\frac{\beta}{\lossSD} \right)}.
    \label{eqn:rd_link_mrc}
\end{align}
In the last equality of~\eqref{eqn:rd_link_mrc} we assume that $\lossRD{k} \neq \lossSD$.

Equation~\eqref{eqn:rd_link_mrc} can be extended to support an arbitrary number $K$ of available relays, obtaining
\addtocounter{equation}{2}
\begin{align}
    &\prob \left[ \chSD \lossSD \! + \! \chRD{1} \lossRD{1} > \! \beta, \dots, \chSD \lossSD \! + \! \chRD{K} \lossRD{K} > \! \beta \right] \notag \\
    &= \frac{1}{1 - \sum_{k=1}^K \frac{\lossSD}{\lossRD{k}}} \prod_{k=1}^K \eul ^ { -\frac{\beta}{\lossRD{k}} } + \left(1 - \frac{1}{1 - \sum_{k=1}^K \frac{\lossSD}{\lossRD{k}}} \right) \eul ^ { -\frac{\beta}{\lossSD} } \notag \\
    &= \coopgain \prod_{k=1}^K \eul ^ {-\frac{\beta}{\lossRD{k}}} + \left(1 - \coopgain \right) \eul ^ {-\frac{\beta}{\lossSD}},
    \label{eqn:multiple_rd_link_mrc_closed}
\end{align}
where the quantity $\coopgain$ is defined as $\coopgain = \left( 1 - \sum_{k=1}^K \frac{\lossSD}{\lossRD{k}} \right) ^ {-1}$. We again assume that $\coopgain \neq 1$.

We can now calculate $P[A]$ as shown in~\eqref{eqn:coop_link_mrc}. Note that in~\eqref{eqn:coop_link_mrc} $(a)$ follows from the assumption that the fading coefficient of the link connecting an interferer $u$ to $\destination$ remains constant over the two consecutive time slots, and so the SIR is given by~\eqref{eqn:sir_mrc}\footnote{A result similar to the one of~\eqref{eqn:coop_link_mrc} in the case where the fading coefficients of the two slots are independent and for $N = 2$ can be obtained according to Theorem~1 in~\cite{tanbourgi14a}.}. Also, $(b)$ follows from substituting $\beta = \thr \intD$ in~\eqref{eqn:multiple_rd_link_mrc_closed} and by multiplying the two terms by the probability $\prod_{k=1}^K \mathrm{e} ^ { \left(- \thrSR{k} \sum_{u \in \ppp} \chuR{k} \lossuR{k} \ualoha \right) }$ that the $K$ source-relay links are successful. The rest of the steps are similar to those used in deriving~\eqref{eqn:dir_link_sel}.

\subsection{Results for the One Relay Case}
\label{sec:onerelay}
We provide simplified expressions and numerical results for the case where a single relay $\relay{1}$ is available. When the destination adopts SC, the SP becomes
\begin{equation}
    \Psucc = \prob \left[ \Esd \cup \Esrd{1} \right] = \prob \left[ \Esd \right] + \prob \left[ \Esrd{1} \right] - \prob \left[ \Esd \cap \Esrd{1} \right].
    \label{eqn:succ_onerelay_sc}
\end{equation}
The first term in~\eqref{eqn:succ_onerelay_sc} can be found by substituting $K = 0$ in~\eqref{eqn:dir_link_sel}. We obtain
\begin{align}
    \prob[\Esd] &= \prob\left[ \chSD \lossSD > \thr \intD \right] \notag \\
    &= \exp \left( -\dens \int_{\mathbb{R}^2} \left[ 1 - \left( \frac{\txp}{1 + \thrSD \lossxD} + 1 - p \right) \right] \dd x \right).
    \label{eqn:dir_link_onerelay}
\end{align}
The second and third terms in~\eqref{eqn:succ_onerelay_sc} are instead calculated by substituting $K = 1$ in~\eqref{eqn:coop_link_sel} and~\eqref{eqn:dir_link_sel}, respectively, obtaining
\begin{align}
    \prob[\Esrd{1}] &= \prob\left[ \chSR{1} \lossSR{1} > \thr \intR{1}, \chRD{1} \lossRD{1} > \thr \intD \right] \notag \\ 
    &= \exp \left(-\dens \int_{\mathbb{R}^2} \left[ 1 - \left( \frac{\txp}{\left( 1 + \thrRD{1} \lossxD \right)} \cdot \right. \right. \right. \notag \\ &\left. \left. \left. \frac{1}{\left(1 + \thrSR{1} \lossxR{1} \right)} + 1 - p \right) \right] \dd x \right)
    \label{eqn:relay_link_onerelay_sc}
\end{align}
and
\begin{align}
    \prob[\Esd \cap \Esrd{1}] &= \prob\left[ \chSD \lossSD > \thr \intD, \chSR{1} \lossSR{1} > \thr \intR{1}, \chRD{1} \lossRD{1} > \thr \intD \right] \notag \displaybreak[0] \\
    &= \exp \left( -\dens \int_{\mathbb{R}^2} \left[ 1 - \left( \frac{\txp}{\left( 1 + \left( \thrSD + \thrRD{1} \right) \lossxD \right)} \cdot \right. \right. \right. \notag \\ &\left. \left. \left. \frac{1}{\left(1 + \thrSR{1} \lossxR{1} \right)} + 1 - p \right) \right] \dd x \right).
    \label{eqn:coop_link_onerelay_sc}
\end{align}

When the destination adopts MRC, the SP is
\begin{equation}
    \Psucc = \prob \left[ \Esd \cup \EsrdMRC{1} \right] = \prob \left[ \Esd \right] + \prob \left[ \EsrdMRC{1} \right] - \prob \left[ \Esd \cap \EsrdMRC{1} \right].
    \label{eqn:succ_onerelay_mrc}
\end{equation}
The first term in~\eqref{eqn:succ_onerelay_mrc} is given in~\eqref{eqn:dir_link_onerelay}. The second and third terms in~\eqref{eqn:succ_onerelay_mrc} can be derived by substituting $K = 1$ in~\eqref{eqn:coop_link_mrc} and~\eqref{eqn:dir_link_mrc}, respectively, and are shown in~\eqref{eqn:relay_link_onerelay_mrc} and \eqref{eqn:coop_link_onerelay_mrc}.

\begin{figure}[!t]
    \centering
        \input{Figures/dir-vs-single-relay-psucc}
    \caption{SP $\Psucc$ when varying $\txp$ for different values of $\dens$ for a non-cooperative scenario ($\dens = 0.2$:~\ref{plt:dir-vs-single-dir-lam-0dot2}, $\dens = 0.6$:~\ref{plt:dir-vs-single-dir-lam-0dot6}, $\dens = 1$:~\ref{plt:dir-vs-single-dir-lam-1dot0}, $\dens = 2$:~~\ref{plt:dir-vs-single-dir-lam-2dot0}) and for a single-relay, SC scenario ($\dens = 0.2$:~\ref{plt:dir-vs-single-sel-lam-0dot2}, $\dens = 0.6$:~\ref{plt:dir-vs-single-sel-lam-0dot6}, $\dens = 1$:~\ref{plt:dir-vs-single-sel-lam-1dot0}, $\dens = 2$:~\ref{plt:dir-vs-single-sel-lam-2dot0}). We set $\thr = 1$.}
    \label{fig:direct_vs_single_succ}
\end{figure}

In Figs.~\ref{fig:direct_vs_single_succ} and \ref{fig:single_relay_succ} we present numerical results for a setting where the source is located at $(0,0)$ and the destination at $(1,0)$. We assume $\thr = 1$. In the figures, we plot the SP $\Psucc$ as a function of the interferer transmission probability $\txp$, for $\dens=0.2$, $0.6$, $1$, and $2$. 

\begin{figure}[!t]
    \centering
        \input{Figures/single-relay-psucc}
    \caption{SP $\Psucc$ for a scenario with a single relay when varying the transmission probability $\txp$, for different values of $\dens$, and adopting SC ($\dens = 0.2$:~\ref{plt:single-sel-lam-0dot2}, $\dens = 0.6$:~\ref{plt:single-sel-lam-0dot6}, $\dens = 1$:~\ref{plt:single-sel-lam-1dot0}, $\dens = 2$:~\ref{plt:single-sel-lam-2dot0}) and MRC ($\dens = 0.2$:~\ref{plt:single-mrc-lam-0dot2}, $\dens = 0.6$:~\ref{plt:single-mrc-lam-0dot6}, $\dens = 1$:~\ref{plt:single-mrc-lam-1dot0}, $\dens = 2$:~\ref{plt:single-mrc-lam-2dot0}). We set $\thr = 1$.}
    \label{fig:single_relay_succ}
\end{figure}

In particular, Fig.~\ref{fig:direct_vs_single_succ} compares the SP of the system with no relays with that of a cooperative system where a single relay, located at $(0.25,0)$, is available, and SC is adopted at the destination. In order to have a fair comparison, the transmission power is doubled when there is no relay. The figure reveals that cooperation significantly increases the SP, compared to the case where no relays are available. This is due to the well-understood fact that when the direct $\source-\destination$ link is not available due to a deep fade, it is possible that both the $\source - \relay{1}$ and $\relay{1} - \destination$ links are not in a deep fade and thus may be able to support the communication. This important result is also verified in our setting.

Fig.~\ref{fig:single_relay_succ} compares the SP for the two detection strategies. Again, the single relay is located at $(0.25,0)$. One can see that adopting MRC at the destination provides a higher value of $\Psucc$, hence this combining technique improves the system performance. However, the improvement is only modest, and as MRC is harder to implement, designers should take note. A more detailed analysis of the benefits provided by adopting MRC at the destination will be presented in the following subsection, for the case of multiple relays.

\subsection{Results for the Multiple Relays Case}

We now provide numerical results for settings where multiple relays are available. Again, we consider networks consisting of a source located at $(0,0)$ and a destination located at $(1,0)$. The rest of the parameters, including the number of available relays and their positions, will be specified for each considered case individually. 

Fig.~\ref{fig:sel_vs_mrc} shows the SP, for both SC and MRC, when $N=1$, $3$, or $5$ relays are clustered together at position $(\cluster, 0)$ and move along the line connecting $\source$ and $\destination$. Therefore, all relays share the same distance-dependent path loss values toward $\source$ and $\destination$, however fading values on links connecting different relays to $\source$ and $\destination$ are independent. We plot the SP $\Psucc$ versus the location of the relay cluster $\cluster$ for two different communication scenarios, namely a `good scenario' with parameters $\thr = 0.1$, $\dens = 0.5$ and $\txp = 1$, and a `harsh scenario' with parameters $\thr = 1$, $\dens = 1$ and $\txp = 1$.

\begin{figure}[!t]
    \centering
        \input{Figures/sel-vs-mrc}
    \caption{SP $\Psucc$ when varying the number $N$ of available relays and adopting SC ($N=1$: \ref{plt:sel-vs-mrc-sel-good-N-1}, $N=3$: \ref{plt:sel-vs-mrc-sel-good-N-3}, $N=5$: \ref{plt:sel-vs-mrc-sel-good-N-5}) and MRC ($N=1$: \ref{plt:sel-vs-mrc-mrc-good-N-1}, $N=3$: \ref{plt:sel-vs-mrc-mrc-good-N-3}, $N=5$: \ref{plt:sel-vs-mrc-mrc-good-N-5}).  Relays are placed at $(\cluster, 0)$. Thin lines are for a `good scenario' ($\thr = 0.1$, $\dens = 0.5$ and $\txp = 1$), and thick lines are for a `harsh scenario' ($\thr = 1$, $\dens = 1$ and $\txp = 1$).}
    \label{fig:sel_vs_mrc}
\end{figure}

As expected, $\Psucc$ increases when more relays are available, but the increase is more pronounced in the harsh scenario. Furthermore, it is best to place relays approximately in the middle of the $\source-\destination$ line, but a bit closer to $\destination$. Finally, the improved decoding performance provided by the MRC decoder results in a higher $\Psucc$, but this result is only pronounced when the relays are close to the source, and the benefits vanish when the relays move close to the destination. Indeed, when the relay has received the packet and is close to the destination, the power received at the destination from the source is very small compared to the power received from the relay. Adding these powers, according to MRC, will not be much better than operating with SC, which only employs the (much) larger one of the two powers. On the other hand, when the used relay is close to the source, the two powers are similar, and hence MRC has a notable advantage. To conclude, designers should try to place relays at around the middle between  source and destination and with a bias towards the destination, should employ numerous relays only when the system operates in harsh environments, and should use MRC only when the relays are placed close to the source.

\begin{figure}[!t]
    \centering
        \input{Figures/random-pos}
    \caption{SP $\Psucc$ when adopting MRC and varying the number $N$ of available relays ($N=1$: \ref{plt:random-mrc-good-N-1}, $N=2$: \ref{plt:random-mrc-good-N-2}, $N=3$: \ref{plt:random-mrc-good-N-3}, $N=4$: \ref{plt:random-mrc-good-N-4}, $N=5$: \ref{plt:random-mrc-good-N-5}). Relays are placed randomly in a square of side length $\edge$ centered in the middle between $\source$ and $\destination$. We plot the average SP versus $\edge$ over $200$ random relay positions. Thin lines are for a `good scenario' with $\thr = 0.1$, $\dens = 0.5$ and $\txp = 1$, and thick lines are for a `harsh scenario' with $\thr = 1$, $\dens = 1$ and $\txp = 1$.}
\label{fig:random_pos}
\end{figure}
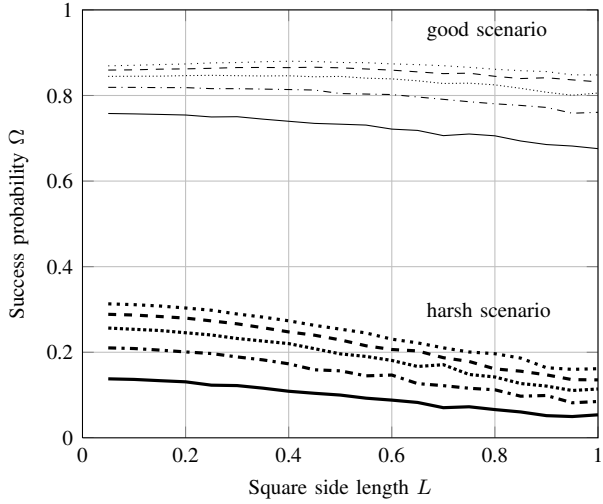

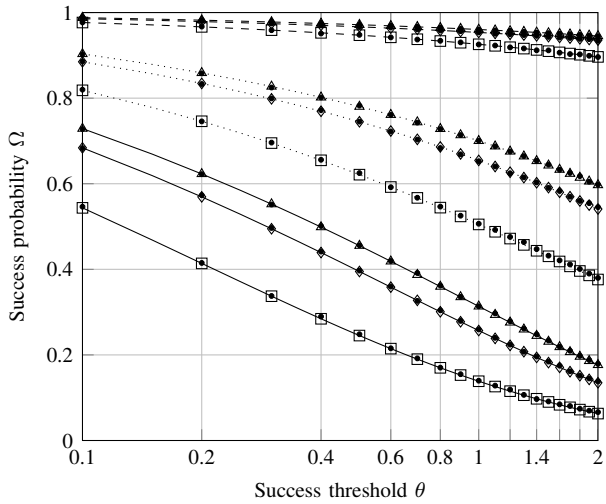
\begin{figure}[!t]
    \centering
        \input{Figures/succ-vs-th}
    \caption{SP $\Psucc$ as a function of the success threshold $\thr$  when adopting MRC and varying the number $N$ of available relays for $\dens = 0.1$ and $\txp = 0.5$ ($N=1$: \ref{plt:succ-vs-th-lambda-0dot1-p-0dot5-N-1}, $N=3$: \ref{plt:succ-vs-th-lambda-0dot1-p-0dot5-N-3}, $N=5$: \ref{plt:succ-vs-th-lambda-0dot1-p-0dot5-N-5}), for $\dens = 0.5$ and $\txp = 0.75$ ($N=1$: \ref{plt:succ-vs-th-lambda-0dot5-p-0dot75-N-1}, $N=3$: \ref{plt:succ-vs-th-lambda-0dot5-p-0dot75-N-3}, $N=5$: \ref{plt:succ-vs-th-lambda-0dot5-p-0dot75-N-5}), and for $\dens =1$ and $\txp = 1$ ($N=1$: \ref{plt:succ-vs-th-lambda-1-p-1-N-1}, $N=3$: \ref{plt:succ-vs-th-lambda-1-p-1-N-3}, $N=5$: \ref{plt:succ-vs-th-lambda-1-p-1-N-5}). Relays are placed in the middle between $\source$ and $\destination$. Solid dots~(\ref{plt:sim-relay-mrc-lambda-0dot1-p-0dot5-N-1}) are obtained through simulations by averaging $50,000$ transmission attempts.} 
 \label{fig:succ-vs-thresh}
\end{figure}

 \begin{figure}[!t]
    \centering
        \input{Figures/psucc-times-rate-vs-th}
    \caption{The product $\Psucc \ln(1 + \thr)$ as a function of $\thr$  when adopting MRC, for $\dens=0.1$ and $\txp=0.5$ ($N = 1$:~\ref{plt:relay-mrc-const-lambda-0dot1-p-0dot5-N-1}, $N = 3$:~\ref{plt:relay-mrc-const-lambda-0dot1-p-0dot5-N-3}, $N = 5$:~\ref{plt:relay-mrc-const-lambda-0dot1-p-0dot5-N-5}),  $\dens=0.25$ and $\txp=0.75$ ($N = 1$:~\ref{plt:relay-mrc-const-lambda-0dot25-p-0dot75-N-1}, $N = 3$:~\ref{plt:relay-mrc-const-lambda-0dot25-p-0dot75-N-3}, $N = 5$:~\ref{plt:relay-mrc-const-lambda-0dot25-p-0dot75-N-5}), $\dens=0.75$ and $\txp=0.5$ ($N = 1$:~\ref{plt:relay-mrc-const-lambda-0dot75-p-0dot5-N-1}, $N = 3$:~\ref{plt:relay-mrc-const-lambda-0dot75-p-0dot5-N-3}, $N = 5$:~\ref{plt:relay-mrc-const-lambda-0dot75-p-0dot5-N-5}), and $\dens=1$ and $\txp=1$ ($N = 1$:~\ref{plt:relay-mrc-const-lambda-1-p-1-N-1}, $N = 3$:~\ref{plt:relay-mrc-const-lambda-1-p-1-N-3}, $N = 5$:~\ref{plt:relay-mrc-const-lambda-1-p-1-N-5}). Relays are placed in the middle between $\source$ and $\destination$.}
     \label{fig:throughput}
 \end{figure}

In Fig.~\ref{fig:random_pos} we investigate further the effects of relay placement, for the two scenarios of Fig.~\ref{fig:sel_vs_mrc} and MRC, but for a different relay placement scheme. In particular, $N$ relays, for $N \in \{1,2,3,4,5\}$, are uniformly and independently placed in a square of side length $\edge$ centered in the middle point between $\source$ and $\destination$. We plot the SP $\Psucc$, averaged over $200$ random relay placements, versus the side length $\edge$.

As shown in the figure, in the good scenario, $\Psucc$ hardly changes when the side length increases, i.e., moving from a situation where relays are very close to the middle point between the source and the destination to a situation where they are spread over a larger area. This is explained by noting that, when having multiple relays, it is still likely that one is located in a beneficial position, and this relay will be used irrespective of the locations of the rest. Furthermore, the case where all relays are very close to the middle point between $\source$ and $\destination$ exhibits high spatial correlation properties, hence reducing the benefit of the smaller path loss. On the other hand, when operating in bad communication conditions, increasing the side length yields a notably lower $\Psucc$, which means that, in this case, the benefits of having small path losses outweigh the benefits of having decorrelated interference  at the relays. In both scenarios, however, it turns out that overall it is best for the relays to be placed in a way that the effects of path loss are minimized, even if this means that the interference powers experienced by the relays are correlated (due to the fact that relays are placed close to each other).

Fig.~\ref{fig:succ-vs-thresh} shows the effect  of changing the decoding success threshold $\thr$ in the case of MRC. We plot the SP $\Psucc$ versus the threshold $\thr$ for $N=1,3$, and $5$. Results are plotted for $\dens=0.1$ and $\txp=0.5$, for $\dens=0.5$ and $\txp=0.75$, and for $\dens=1$ and $\txp=1$. Relays are placed in the middle between $\source$ and $\destination$.

As expected, adopting multiple relays increases the SP. Also, when the density of interferers $\dens$ and their probability of accessing the channel $\txp$ are low, the SP is very high for a wide range of thresholds $\thr$. On the other hand, when the interference power at the destination increases (i.e., $\dens$ and $\txp$ increase), the SP is significantly reduced if the threshold is set at a high value.

Finally, as a means of double-checking the correctness of these analytical results, we also obtain all data points of Fig.~\ref{fig:succ-vs-thresh} using simulations. Each data point is calculated by taking the average over $50,000$ transmission attempts. Note that analysis and simulations match very well in all cases. 

We conclude our numerical investigation in Fig.~\ref{fig:throughput} where we plot the product $\Psucc \ln(1 + \thr)$ as a function of the success threshold $\thr$, for the case where MRC is adopted and for $N=1$, $3$ and $5$. Results are plotted for $\dens=0.1$ and $\txp=0.5$, for $\dens=0.25$ and $\txp=0.75$, for $\dens=0.75$ and $\txp=0.5$, and for $\dens=1$ and $\txp=1$. Again, relays are placed in the middle between $\source$ and $\destination$.

Note that $\ln(1 + \thr)$ is the Shannon bound (in nats per $\mathrm{Hz}$) for the data rate if the Signal to Noise Ratio (SNR) equals $\thr$~\cite{cover1}. Therefore, $\ln(1 + \thr)$  is the maximum theoretical rate of communication per unit of spectrum in our setting if no rate adaptation is employed, the effects of interference equal the effects of noise, and if, as we have assumed, the receiver is required to decode the packet successfully whenever the SIR exceeds $\thr$. Given that $\Psucc$ is the probability that a single transmission attempt is successful, the product $\Psucc \ln(1 + \thr)$ is the expected volume of data transmitted per transmission attempt, in nats per $\mathrm{Hz}$, if each time slot is of unit duration. Therefore, this product is a measure of throughput. Observe that the optimal value of $\theta$, which maximizes the product, decreases when the harshness of the scenario increases and increases when $N$ increases. 

\section{Multiple Transmission Attempts}
\label{sec:arq}
We now assume that the wireless system employs the following simple ARQ scheme: whenever a transmission attempt fails, the source initiates another one. Let $T$ be the discrete random variable describing the number of transmission attempts until one attempt is successful. In this section, we calculate the distribution of $T$ and the throughput of a system based on this ARQ scheme, under the two receiver models (SC and MRC) and two mobility models, the Stationary Interferer Model (SIM) and the Meteoric Interferer Model (MIM).  We also present simulation results for the third mobility model, the Traveling Interferer Model (TIM), that bridges the gap between SIM and MIM. We conclude the section by discussing a tradeoff that emerges between the energy efficiency and throughput of an alternative, opportunistic scheme. 

\subsection{Distribution of $T$}
\label{sub:distribution_T}

The distribution of $T$ can be found in a straightforward manner in the case of MIM. Indeed, both the interferer locations and the fading coefficients of distinct transmission attempts are independent; therefore, the events that distinct transmission attempts are successful are also independent. Hence, the number of attempts $T$ follows the geometric distribution with parameter $\Psucc$ (i.e., the probability of success of the consecutive experiments); its probability mass function (pmf) is
\begin{align}
    \PsExactT{t} \triangleq \prob [T=t] = \left(1 - \Psucc \right) ^{T-1} \Psucc, \quad t=1,2,\ldots
    \label{eqn:succ_arq_hmim}
\end{align}
with $\Psucc$ given by~\eqref{eqn:succ_union} for SC or by~\eqref{eqn:succ_union_mrc} for MRC.

Handling SIM is not as easy, due to the fact that different transmission attempts are correlated through the common placement of interferers. To calculate the pmf of $T$ in this case, let $\evT{i}$ denote the event that the $i$-th transmission attempt is a success, $i=1,2,\dots$. The retransmission scheme succeeds exactly at the $t$-th attempt when all the previous $(t-1)$ attempts are unsuccessful, and attempt $t$ is successful. Hence, 
\begin{align}
 \PsExactT{t} \triangleq	\prob [T=t] = \prob \left[ \left( \bigcap_{\tau=1}^{t-1} \cevT{\tau}  \right) \cap \evT{t} \right], \quad t=1,2,\ldots
    \label{eqn:succ_arq}
\end{align}

Using this starting point, we have
\begin{align}
\PsExactT{t} &= \prob[T=t] = \prob \left[ \bigcap_{\tau=1}^{t-1} \cevT{\tau} \cap \evT{t} \right] \notag \\
    &\overset{(a)}{=} \expect_{\ppp} \left[ \prob \left[ \bigcap_{\tau=1}^{t-1} \cevT{\tau} \cap \evT{t} | \ppp \right] \right] \notag \\
    &\overset{(b)}{=} \expect_{\ppp} \left[ \left( \prod_{\tau=1}^{t-1} \left(1 - \prob \left[\evT{\tau} | \ppp \right] \right) \right) \prob \left[\evT{t} | \ppp \right] \right] \notag \\
    &\overset{(c)}{=} \expect_{\ppp} \left[ \left( \prod_{\tau=1}^{t-1} \left(1 - \prob \left[\evT{} | \ppp \right] \right) \right) \prob \left[\evT{} | \ppp \right] \right] \notag \\
    &= \expect_{\ppp} \left[ \left(1 - \prob \left[\evT{} | \ppp \right] \right)^{t-1} \prob \left[\evT{} | \ppp \right] \right] \notag \\
    &\overset{(d)}{=} \expect_{\ppp} \left[ \sum_{\tau=0}^{t-1} \binom{t-1}{\tau} (-1)^\tau \left( \prob \left[\evT{} | \ppp \right] \right)^{\tau+1} \right] \notag \\
    &= \sum_{\tau=0}^{t-1} \binom{t-1}{\tau} (-1)^\tau \expect_{\ppp} \left[ \left( \prob \left[\evT{} | \ppp \right] \right)^{\tau+1} \right].
    \label{eqn:succ_arq_sim}
\end{align}
Equation $(a)$ follows from conditioning on a particular realization of the PPP; $(b)$ follows from noting that once the interferer locations are fixed, the events that different attempts are successful are independent, since the fading powers and the behavior of interferers at different times are independent; in $(c)$ we remove the index $\tau$ from the success events, since the probability of the event $\evT{\tau}$ does not depend on the time slot $\tau$; finally, $(d)$ is obtained by applying the binomial power expansion. 

To obtain an expression for the quantity $\left( \prob \left[\evT{} | \ppp \right] \right)^{\tau+1}$ of~\eqref{eqn:succ_arq_sim} we first note that, by the inclusion-exclusion principle,
\begin{equation}
    S = \Esd \cup \left( \bigcup_{n=1}^N \Esrdplain{n} \right) \Rightarrow P[S] = \sum_{A \in \powerset{\Esucc}} (-1)^{\card{A} + 1} \prob[A],
    \label{eqn:succ_union2}
\end{equation}
where $\Esrdplain{n}=\Esrd{n}$ in the case of SC and $\Esrdplain{n}=\EsrdMRC{n}$ in the case of MRC, and $\Esucc = \{ \Esd, \Esrdplain{1}, \dots, \Esrdplain{N} \}$. This equation also holds when there is conditioning, therefore
\begin{align}
    \left( \prob \left[\evT{} | \ppp \right] \right)^{\tau+1} &= \left( \sum_{A \in \powerset{\Esucc}} (-1)^{\card{A} + 1} \prob[A | \ppp ] \right)^{\tau+1} \notag \\
    &\overset{(a)}{=} \sum_{a_1 + \dots + a_{|\powerset{\Esucc}|} = \tau + 1} \left( \frac{(\tau+1)!}{a_1! \cdot \dots \cdot a_{|\powerset{\Esucc}|}!} \right. \cdot \notag \\ &\left. \prod_{i=1}^{|\powerset{\Esucc}|}\left( (-1)^{\card{A_i} + 1} \prob[A_i | \ppp] \right)^{a_i} \right),
    \label{eqn:succ_arq_part}
\end{align}
where $(a)$ follows from the multinomial theorem and from denoting the $i$-th element of $\powerset{\Esucc}$ with $A_i$. Therefore, it suffices to find expressions for $\prob[A_i | \ppp]$.

When SC is employed and $A = \{ \Esd, \Esrd{1}, \dots, \Esrd{K}\}$, we have, following the steps of~\eqref{eqn:dir_link_sel} of Section~\ref{sec:sel},
\begin{align}
    \prob[A | \ppp] &= \prod_{u \in \ppp} \left( \!\! \left( \frac{\txp \prod_{k=1}^K \frac{1}{1 + \thrSR{k} \lossuR{k}} }{1 + \left( \thrSD + \sum \limits_{k=1}^K \thrRD{k} \right) \lossuD} \right) + 1 - \txp \right).
    \label{eqn:dir_link_sel_ppp}
\end{align}

Similarly, for SC and $A = \{\Esrd{1}, \dots, \Esrd{K}\}$, we follow~\eqref{eqn:coop_link_sel} and obtain
\begin{align}
    \prob[A | \ppp] \! &= \! \prod_{u \in \ppp} \left( \!\! \left( \! \frac{\txp \prod_{k=1}^K \frac{1}{1 + \thrSR{k} \lossuR{k}}}{1 + \sum \limits_{k=1}^K \thrRD{k} \lossuD} \right) + 1 - \txp \right).
    \label{eqn:dir_coop_sel_ppp}
\end{align}

When MRC is employed and $A = \{ \Esd, \EsrdMRC{1}, \dots, \EsrdMRC{K} \}$ we have, according to~\eqref{eqn:dir_link_mrc},
\begin{align}
    \prob[A | \ppp] &= \prod_{u \in \ppp} \left( \left( \frac{\txp}{1 + \thrSD \lossuD} \prod_{k=1}^K \frac{1}{1 + \thrSR{k} \lossuR{k}} \right) + 1 - \txp \right).
    \label{eqn:dir_link_mrc_ppp}
\end{align}

Finally, when MRC is employed and $A = \{\EsrdMRC{1}, \dots, \EsrdMRC{K} \}$ we have, according to~\eqref{eqn:coop_link_mrc},
\begin{align}
    \prob[A | \ppp] &= \coopgain \! \prod_{u \in \ppp} \! \left( \!\! \left( \frac{\txp}{1 \! + \! \sum \limits_{k=1}^K \thrRD{k} \lossuD} \! \prod_{k=1}^K \frac{1}{1 \! + \! \thrSR{k} \lossuR{k}} \right) \! + \! 1 \! - \! \txp \right) \! + \notag \\ &\quad \left(1 \! - \! \coopgain \right) \! \prod_{u \in \ppp} \! \left( \!\! \left( \frac{\txp}{1 \! + \! \thrSD \lossuD} \! \prod_{k=1}^K \frac{1}{1 \! + \! \thrSR{k} \lossuR{k}} \right) \! + \! 1 \! - \! \txp \right) \!.
    \label{eqn:coop_link_mrc_ppp}
\end{align}

We now have the expression~\eqref{eqn:succ_arq_part} for the quantity $\left( \prob \left[\evT{} | \ppp \right] \right)^{\tau+1}$, with all the expressions $\prob[A_i | \ppp]$ appearing in the right hand side of~\eqref{eqn:succ_arq_part} given by~\eqref{eqn:dir_link_sel_ppp},~\eqref{eqn:dir_coop_sel_ppp},~\eqref{eqn:dir_link_mrc_ppp}, and~\eqref{eqn:coop_link_mrc_ppp}. The final result $\prob[T=t]$ is then obtained by substituting the expression~\eqref{eqn:succ_arq_part} we have for $\left( \prob \left[\evT{} | \ppp \right] \right)^{\tau+1}$ in~\eqref{eqn:succ_arq_sim}, exchanging the orders of the sum contained in $\left( \prob \left[\evT{} | \ppp \right] \right)^{\tau+1}$ and the $\expect_{\ppp}$ operator, and finally applying the probability generating functional theorem for $\ppp$~(cf.~\cite{haenggi13:book}, (4.8)).

\subsection{Numerical Results}

\begin{figure}[!t]
    \centering
        \input{Figures/renewal-ind-vs-dep}
    \caption{The cdf $P[T \leq t]$ of the number of attempts $T$ of the ARQ scheme with MRC for the SIM (\ref{plt:renewal-mrc-dep-good}), MIM (\ref{plt:renewal-mrc-ind-good}) and TIM (\ref{plt:sim-renewal-mrc-speed-1-good}) models. In the case of TIM, we simulate interferer speeds $\speed = 0.1, 0.2, 0.5, 0.75, 1$, and $2$ units of length per inter-attempt time $D_0$. The source $\source$ is located at $(0,0)$, the destination $\destination$ is located at $(1,0)$, and $N = 3$ relays are clustered in $(0.5,0)$. Thin lines are for a `good scenario' ($\thr = 0.1$, $\dens = 0.5$ and $\txp = 1$), and thick lines are for a `harsh scenario' ($\thr = 1$, $\dens = 1$ and $\txp = 1$).}
    \label{fig:renewal_ind_vs_dep}
\end{figure}
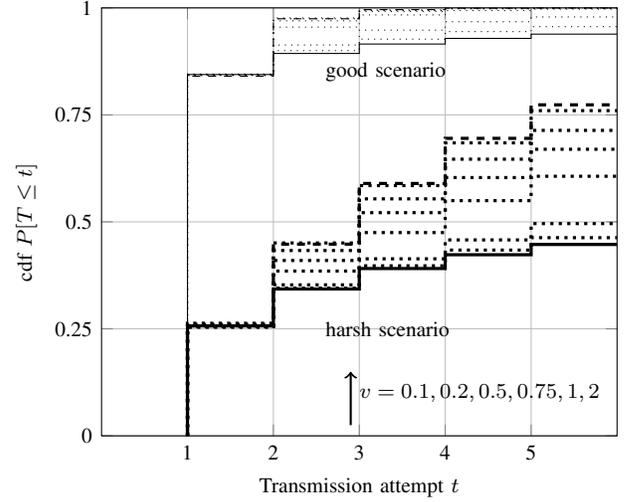

\begin{figure}[!t]
    \centering
        \input{Figures/renewal-num-relays}
    \caption{The cdf $P[T \leq t]$ of the number of attempts $T$ of the ARQ scheme without relays (\ref{plt:renewal-mrc-good-N-0}) and with relays ($N = 1$: \ref{plt:renewal-mrc-good-N-1}, $N = 2$: \ref{plt:renewal-mrc-good-N-2}, $N = 3$: \ref{plt:renewal-mrc-good-N-3}), under SIM and MRC. Thin lines are for a `good scenario' ($\thr = 0.1$, $\dens = 0.5$ and $\txp = 1$), and thick lines are for a `harsh scenario' ($\thr = 1$, $\dens = 1$ and $\txp = 1$).}
    \label{fig:renewal_coop_vs_dir}
\end{figure}
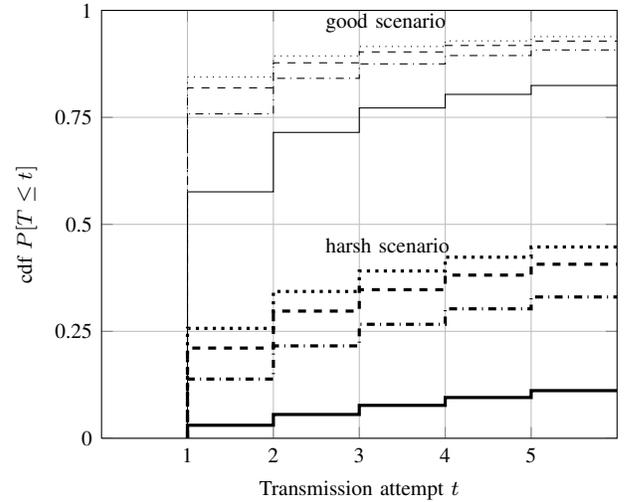

Having calculated the cdf of $T$, we now study how it is affected by the interferer mobility model and the number of relays, using a few selected settings. The source $\source$ is located at $(0,0)$, the destination $\destination$ is located at $(1,0)$, and relays (if they are assumed) are located at $(0.5,0)$. We consider a `good scenario', with $\thr = 0.1$, $\dens = 0.5$ and $\txp = 1$, and a `harsh scenario', with $\thr = 1$, $\dens = 1$ and $\txp = 1$. The cdf adopting SC is very close to the cdf adopting MRC, as one expects from the results of Fig.~\ref{fig:sel_vs_mrc} for the single attempt case. Hence, only the cdf using MRC is shown.

Starting with the effects of the interferer mobility model, in Fig.~\ref{fig:renewal_ind_vs_dep} we plot the cdf $P[T \leq t]$ for all three mobility models, i.e., the Stationary Interferer Model (SIM), the Meteoric Interferer Model (MIM), and the Traveling Interferer Model (TIM). Results for TIM are obtained through Monte Carlo simulations, by calculating the empirical cdf of $T$ for $50,000$ transmission attempts, and for interferer speeds  $\speed = 0.1, 0.2, 0.5, 0.75, 1, 2$ units of length per inter-attempt time $D_0$. We assume $N = 3$ relays. 

Two important observations are in order. Firstly, adopting MIM leads to an increased performance, with respect to SIM. Indeed, additional transmission attempts provide less benefit in the case of SIM, because interference levels at different transmission attempts are correlated. Therefore, if some attempts have already failed, there must be many interfering transmissions nearby, and for this reason subsequent retransmission attempts will very likely also fail. On the other hand, retransmission attempts provide significant benefit in the case of MIM, because in this case the interference levels during different transmission attempts are independent.

Secondly, observe that TIM, which models levels of mobility between those of SIM and MIM, indeed has a performance between the ones of SIM and MIM. For small values of $\speed$, TIM is very close to SIM, whereas for large values of $\speed$, its performance is very close to that of MIM. This suggests that we can use SIM and MIM to upper and lower bound, respectively, the performance of realistic mobility models.

Designers should take the above findings into account; they suggest that in the presence of moving interferers, there might be a tradeoff between the average number of attempts $\expect [T]$ until the transmission is successful (which is proportional to the energy dissipated per packet) and the duration of time $D_0$ between transmission attempts (which is related to throughput). Indeed, if we increase $D_0$ (thus decreasing the throughput, because fewer attempts are made per unit of time) the performance of the system approaches the performance of MIM (because the interferer positions at consecutive attempts are less correlated) and thus, on the average, fewer attempts will be needed per packet. Therefore, reducing the throughput could increase the energy efficiency. As this issue has an obvious practical significance for designers, we explore it in more detail in Section \ref{sub:tradeoff}.

Moving on to the effects of the number of relays, Fig.~\ref{fig:renewal_coop_vs_dir} shows the cdf $P[T \leq t]$ of $T$ for $N=1$, $2$, and $3$, assuming the SIM mobility model, and compares it with the cdf of a non-cooperative system with no relays, but (for reasons of fairness) with double the transmission power. 

The figure reveals that cooperation significantly improves the performance. In particular, there is an almost constant gap (with respect to $t$) between the cdf of the cooperative system and the cdf of the non-cooperative system. Improvements are especially pronounced in the harsh scenario, where the non-cooperative system has very low chances of correctly delivering the packet even after multiple transmission attempts. Therefore, the advantages of cooperative relaying extend also to cases where ARQ schemes are employed.

\subsection{Throughput of ARQ Scheme}
\label{sub:throughput_arq}

We now calculate the throughput $\throughput$ of the ARQ scheme under study, defined here as the average amount of packets successfully delivered per transmission attempt, for the SIM and MIM models. 
We add the realistic constraint that the maximum number of transmission attempts allowed is $T_{\mathrm{max}}$; after $T_\mathrm{max}$ failed transmission attempts the packet is discarded. Furthermore, whenever a packet is discarded or successfully transmitted, the source backs off, i.e., we assume that in the first attempt of a new packet the placement of interferers is independent from their placement during the last transmission attempt of the previous packet, even in the case of SIM.

To calculate the throughput $\throughput$, we recall that the $t$-th transmission attempt succeeds with probability $\PsExactT{t}$.  We note that a packet is successfully delivered if any of the first $T_\mathrm{max}$ transmission attempts succeeds; this event has probability $\sum_{t=1}^{T_\mathrm{max}} \PsExactT{t}$. The average number of attempts equals $\expect [T] = \sum_{t=1} ^ {T_\mathrm{max}} \left(t \PsExactT{t}\right) + T_\mathrm{max} \left(1 - \sum_{t=1} ^ {T_\mathrm{max}} \PsExactT{t}\right)$, where the last term accounts for the $T_\mathrm{max}$ transmission attempts taken by packets that are discarded. The throughput can then be obtained as
\begin{equation}
    \throughput = \frac{\sum_{t=1}^{T_\mathrm{max}} \PsExactT{t}}{\sum_{t=1} ^ {T_\mathrm{max}} (t \PsExactT{t}) + T_\mathrm{max} \left(1 - \sum_{t=1} ^ {T_\mathrm{max}} \PsExactT{t}\right)}.
    \label{eqn:throughput_arq}
\end{equation}

In Fig.~\ref{fig:renewal_vs_th} we plot the throughput $\Lambda$ versus the success threshold $\thr$ for the three mobility models and MRC. In the case of TIM, we assume interferer speeds $\speed = 0.1, 0.2, 0.5, 0.75, 1$, and $2$ units of length per inter-attempt time $D_0$. The source $\source$ is located at $(0,0)$, the destination $\destination$ is located at $(1,0)$, and the single relay is located at $(0.5,0)$. We consider a `good scenario' ($\dens = 0.25$ and $\txp = 1$), and a `harsh scenario' ($\dens = 1$ and $\txp = 1$). Results for SIM and MIM are arrived at by applying (\ref{eqn:throughput_arq}); however, for reasons of verification we also plot simulation results arrived at by simulating $50,000$ transmission attempts. Results for TIM are found exclusively by simulation, again by simulating $50,000$ transmission attempts.

Again, it can be seen that mobility helps to increase the system performance, since transmissions at different slots have a higher probability of being successful, due to the decorrelation of the interference power at the destination. Indeed, when interferers are static, it is very likely that a failed transmission attempt will be followed by another failed transmission attempt, since an interferer that is located in the vicinity of the relays and destination will remain there. This effect wears off as mobility increases. Finally, again we see that the performance of TIM goes from being similar to the performance of SIM for small speeds $v$, to being similar to the performance of MIM for large speeds $v$. 

\begin{figure}[!t]
    \centering
        \input{Figures/renewal-throughput-th}
    \caption{Throughput $\Lambda$ of the ARQ scheme with MRC for SIM (\ref{plt:throughput-vs-th-const-ppp-mrc-lambda-0dot25-p-1-N-1}), MIM (\ref{plt:throughput-vs-th-ind-ppp-mrc-lambda-0dot25-p-1-N-1}) and TIM (\ref{plt:sim-throughput-vs-th-corr-ppp-mrc-lambda-0dot25-p-1-N-1-speed-0dot1}) with interferer speeds $\speed = 0.1, 0.2, 0.5, 0.75, 1, 2$ units of length per inter-attempt time $D_0$. The source $\source$ is located at $(0,0)$, the destination $\destination$ is located at $(1,0)$, and the single relay is located at $(0.5,0)$. Thin lines are for a `good scenario' ($\dens = 0.25$ and $\txp = 1$), and thick lines are for a `harsh scenario' ($\dens = 1$ and $\txp = 1$). Solid dots~(\ref{plt:sim-throughput-vs-th-const-ppp-mrc-lambda-1-p-1-N-1}) are obtained through simulations of the SIM and MIM scenarios, by simulating $50,000$ transmission attempts, and are intended for validation purposes.}
    \label{fig:renewal_vs_th}
\end{figure}
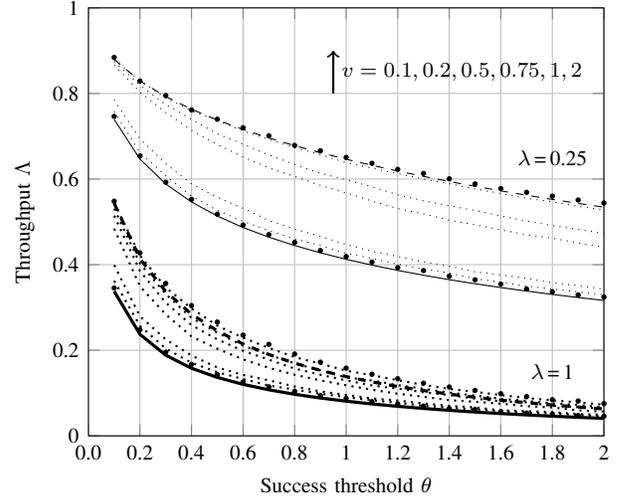

\begin{figure}[!t]
    \centering
        \input{Figures/renewal-throughput-doppler}
    \caption{Throughput $\Lambda$ of the ARQ scheme for SIM, $p=1$, and for different values of $\dens$ and $\thr$. Dashed lines are obtained by simulating a system where fading values on the same link are dependent across retransmissions ($\corr = 0.01$:~\ref{plt:sim-throughput-vs-lam-corr-fad-const-ppp-mrc-th-0dot25-p-1-N-1-fd-0dot01}, $\corr = 0.05$:~\ref{plt:sim-throughput-vs-lam-corr-fad-const-ppp-mrc-th-0dot25-p-1-N-1-fd-0dot05}, $\corr = 0.1$:~\ref{plt:sim-throughput-vs-lam-corr-fad-const-ppp-mrc-th-0dot25-p-1-N-1-fd-0dot1}). Solid lines are analytic results and show the performance of the two limiting cases of having constant (\ref{plt:throughput-vs-lambda-const-fading-const-ppp-th-0dot25-p-1-N-1}) and independent fading coefficients (\ref{plt:throughput-vs-lambda-ind-fading-const-ppp-th-0dot25-p-1-N-1}) across retransmissions. Results are for MRC and for a single relay, located in the middle between $\source$ and $\destination$.}
    \label{fig:renewal_throughput_doppler}
\end{figure}

Finally, in order to evaluate the effects of our assumptions on fading, we compare three different assumptions on the fading coefficients of the same link at different transmission attempts of the same packet, in the case of SIM with the channel access probability $p=1$, a single relay located in the middle between $\source$ and $\destination$, and MRC:
\begin{itemize}
\item These fading coefficients are independent. This has been our assumption until now, and for this case (\ref{eqn:throughput_arq}) applies.
\item These fading coefficients are constant. In this case, the probability that the system is successful at exactly the $t$-th transmission attempt is zero if $t > 1$. This follows from noting that if the fading coefficients and the positions of the interferers are constant across retransmissions, and interferers always transmit, then the interference power and the useful signal power at each receiver remain constant. Hence, either the transmission succeeds at the first transmission attempt, or it will never succeed in the following ones. Therefore, we can still use (\ref{eqn:throughput_arq}) with $P_S^1=\Psucc$ as in the previous case, but with $P_S^t=0$ for all $t \geq 2$. 
\item These fading coefficients are correlated, in particular according to Jakes' model~\cite{jakes1}; the correlation value between the fading coefficients $\channel(i)$, $\channel(j)$ of the same link during transmission attempts $i$ and $j$ is given by $\mathbb{R} \left(\channel(i),\channel(j)\right) = J_0^2 \left(2 \pi f_d |i - j| D_0\right)$, where $J_0$ is the zero-order unmodified Bessel function of the first kind and $f_d$ is the Doppler frequency.
\end{itemize}
Note that the new assumptions only make sense in the context of SIM mobility, as fading coefficients are very sensitive to the placement of nodes. Also note that the fading coefficients of the same link for attempts related to different packets are always independent, and that the fading coefficients of the same link in the two slots of the same transmission attempt are always equal.  

In Fig.~\ref{fig:renewal_throughput_doppler} we plot the throughput $\Lambda$ versus the density $\dens$ for $p=1$, and for the three different assumptions on the fading. In the case of the correlated fading, we consider the cases  $\corr=0.01, 0.05,$ and $0.1$, and arrive at the results by Monte Carlo simulations over $50,000$ transmission attempts. Results for the independent and constant fading coefficient cases are arrived at analytically. We consider the cases $\thr=3$ and $\thr=0.25$.

From Fig.~\ref{fig:renewal_throughput_doppler} one can see that fading correlation across transmission attempts causes a significant throughput decrease. This is ultimately due to the loss of diversity, and mirrors the performance loss seen when interferers are static, compared to the case where they are highly mobile. Hence, Fig.~\ref{fig:renewal_throughput_doppler} confirms that sources of correlation between different transmission are detrimental for the system performance, and it is critical to account for them when designing a cooperative communication system.

\subsection{Tradeoff between Throughput and Energy Efficiency of the Opportunistic Scheme}
\label{sub:tradeoff}

We conclude this section by studying the implications of mobility on the throughput and energy efficiency of our cooperative relaying setting. As we previously noted, one may reduce the rate at which packets are sent (hence reducing the achievable throughput), thereby increasing the probability that attempts that follow unsuccessful attempts are successful themselves (thus eventually increasing the volume of data successfully transmitted per unit of energy).

In order to study this tradeoff, we assume TIM mobility and the following, \emph{opportunistic} transmission scheme, which is different from the ARQ scheme considered until now: The source initiates a transmission attempt. If this attempt is successful, a new packet is sent in the transmission attempt that immediately follows, after the inter-attempt time $D_0$. Otherwise, the source waits for time equal to $\gap$ multiples of the inter-attempt time $D_0$ before starting a new transmission attempt. We refer to $\gap$, which must be a positive integer, as the backoff interval. 

The scheme is very intuitive: when conditions are good, the system keeps transmitting; as soon as there is a failure, the system backs off, waiting for conditions to improve. Clearly, more sophisticated rules could be used, but these exceed the scope of this work. Note that this opportunistic scheme makes no sense under the SIM mobility model, since in that case waiting changes nothing. It also makes no sense under the MIM mobility model, as, in that case, successive transmission attempts are independent irrespective of the value of $\gap$, so it is clearly best to have $\gap=1$.

We assume that one unit of energy is consumed whenever a transmission attempt is carried out, irrespective of the number of relays available, and whether or not the attempt was successful. Clearly, this is a very simple assumption, as in each attempt there may be one or two transmissions; furthermore, there is also energy expended for attempting to receive a packet, and this energy depends on the number of relays available. For this reason, this assumption is suitable for comparing the performance of the same system for different values of $\gap$, which is the purpose of the following discussion, but should not be used for comparing different systems.

We define the throughput $\Lambda_O$ of this opportunistic system  as the number of successful transmission attempts divided by the number of all attempt opportunities, including those during which the system decided to back off. We also define its energy efficiency $\efficiency$ as the number of packets transmitted successfully divided by the total energy expended (i.e., the total number of transmissions). Clearly, we have $\efficiency \leq 1$.

We anticipate that the parameter $\gap$ can be used to trade off throughput with energy efficiency: large values of $\gap$ mean that the scheme takes large breaks, and so the throughput is low; however, transmission attempts attempted after failed ones have good chances of being successful themselves. On the other hand, small values of $\gap$ mean that the breaks are short in duration, and so the throughput is high; however, transmission attempts attempted after failed ones have poor chances of being successful themselves.

The existence of the tradeoff is verified in Fig.~\ref{fig:throughput_burst} which jointly plots the throughput $\Lambda_O$ and the energy efficiency $\efficiency$ for different values of $\gap$. We assume MRC, the TIM model with $\speed=0.1$ units of length per $D_0$, and systems with either $N=1$ or $N=3$ relays located in the middle between $\source$ and $\destination$, and for three scenarios: ($\dens=1, \thr=1$),  ($\dens=0.75, \thr=0.25$), and  ($\dens=0.5, \thr=0.1$). We also set $p=1$.

\begin{figure}[!t]
    \centering
        \input{Figures/throughput-vs-efficiency}
    \caption{Throughput $\Lambda_O$ (thick lines) and energy efficiency $\efficiency$ (thin lines) of the opportunistic scheme for TIM versus the backoff interval $\gap$. Dashed lines are for a single relay, i.e., $N=1$ ($\dens=1,~ \thr=1$:~\ref{plt:throughput-vs-energy-burst-ind-fad-speed-0dot1-lambda-1-p-1-thr-1-N-1}, $\dens=0.75, ~\thr=0.25$:~\ref{plt:throughput-vs-energy-burst-ind-fad-speed-0dot1-lambda-0dot75-p-1-thr-0dot25-N-1}, $\dens=0.5, ~\thr=0.1$:~\ref{plt:throughput-vs-energy-burst-ind-fad-speed-0dot1-lambda-0dot5-p-1-thr-0dot1-N-1}). Solid lines are for $N=3$ ($\dens=1, ~\thr=1$:~\ref{plt:throughput-vs-energy-burst-ind-fad-speed-0dot1-lambda-1-p-1-thr-1-N-3}, $\dens=0.75, ~\thr=0.25$:~\ref{plt:throughput-vs-energy-burst-ind-fad-speed-0dot1-lambda-0dot75-p-1-thr-0dot25-N-3}, $\dens=0.5, ~\thr=0.1$:~\ref{plt:throughput-vs-energy-burst-ind-fad-speed-0dot1-lambda-0dot5-p-1-thr-0dot1-N-3}). Also, $\txp = 1$ and $\speed = 0.1$ units of length per $D_0$. The $N$ relays are clustered at $(0.5,0)$, in the middle between $\source$ and $\destination$.}
    \label{fig:throughput_burst}
\end{figure}

As the figure reveals, increasing $\gap$ leads to a lower throughput, while at the same time the energy efficiency increases. However, after some value of $\gap$, the energy efficiency saturates, while the throughput continues to diminish. Intuitively, there is no gain in waiting for more time than the time needed for the topology to change significantly; after this is achieved, increasing $\gap$ further only reduces the throughout. Therefore, designers employing this scheme, or other, more sophisticated ones, should tune their systems carefully.

\section{Conclusions and Further Work}
\label{sec:conclusions}
In this work we study the performance of a decode-and-forward, cooperative relaying system with multiple relays operating in an interference-limited setting. We present analytical results for two models of interferer placement, both based on spatial Poisson processes and corresponding to two opposite extremes of interferer mobility, namely no mobility and very fast mobility. We also present simulation results for an intermediate mobility model that bridges the gap between them. Two decoding rules are considered: selection combining and maximal ratio combining.

Our analysis provides expressions for the following:
\begin{itemize}
\item the probability of successful transmission in one use of the system, 
\item the cdf of the number of times $T$ the system must be used until there is a success, and
\item the system throughput  assuming a maximum allowed number of retransmissions.
\end{itemize}
These results can be seen as a framework, as well as a building block, for further analysis. They might provide guidelines and insights to the designers of cooperative relaying systems. 

Using the theory of spatial Poisson processes enables us to accurately account for interference dynamics; the price we pay is that expressions are not in closed form and require the numerical calculation of double integrals. However, our techniques and results, apart from being themselves useful, are also useful stepping stones towards deriving other, simpler and/or more general techniques and results. Towards this direction, ongoing and future work involves the formulation of our integrals on the plane through the use of the bi-angular coordinates of~\cite{behnad13a, behnad13b}, with the aim of calculating these integrals in closed form.

Future work also includes the consideration of alternative interferer mobility models, more advanced decoding schemes (such as those involving the use of signals of multiple relays), and realistic channel access strategies (such as those based on carrier sensing).

\bibliographystyle{IEEEtran}
\bibliography{paper}

\vspace*{-1cm}
\begin{biography}
    [{\includegraphics[width=1in,height=1.25in,clip,keepaspectratio]{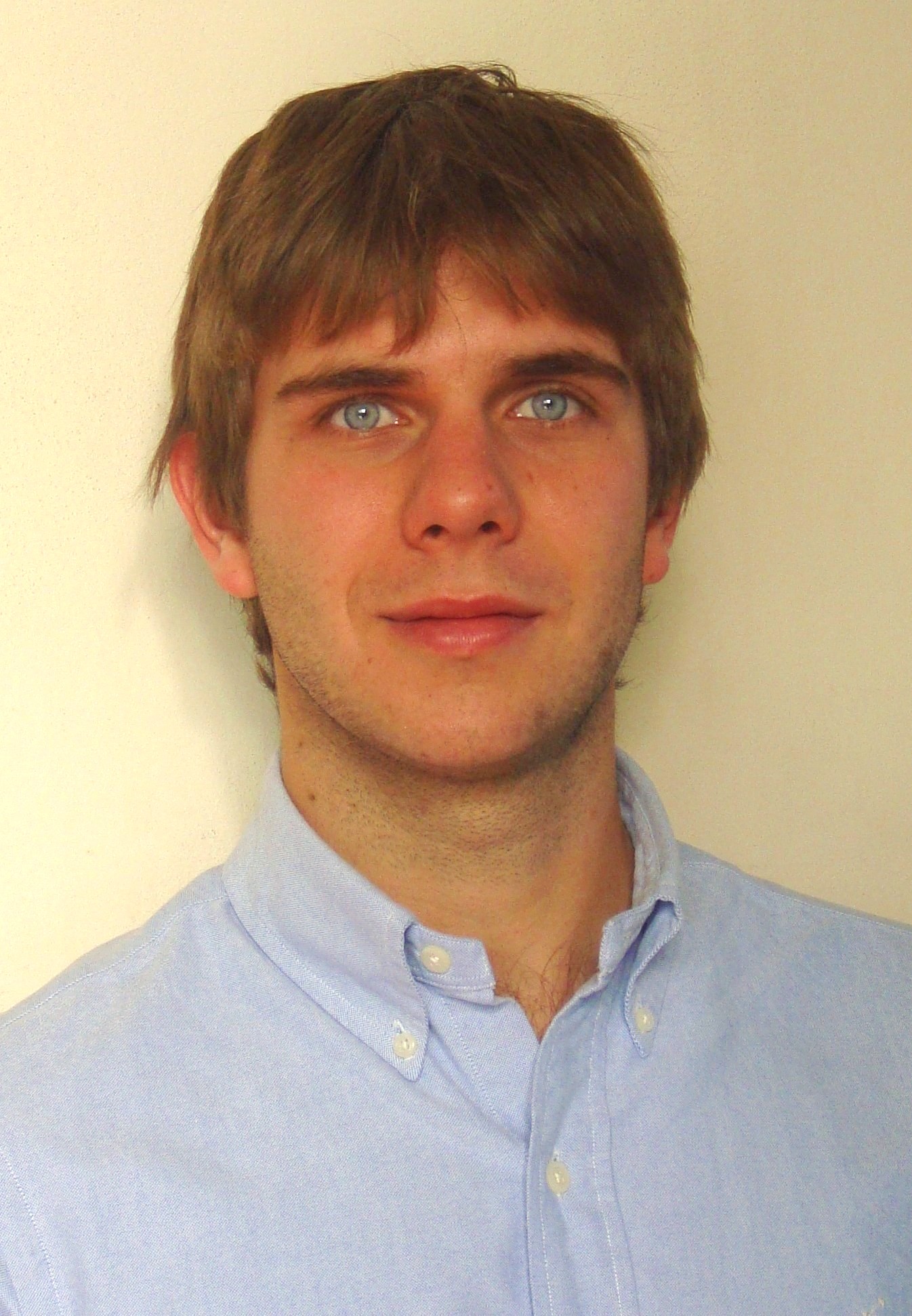}}]{Alessandro Crismani}
Alessandro Crismani received the master degree in telecommunications engineering, from the University of Trieste, Italy, in 2009 (summa cum laude). From June 2009 to November 2009 he was with the Communications Research Group of the University of Southampton, United Kingdom, as a visiting student, and from January 2010 to December 2012 he was with the Telecommunication Group of the University of Trieste, where he worked towards his Ph.D. In 2013 he was with the Mobile Group at the Networked and Embedded Systems of the University of Klagenfurt as a researcher. From January 2014 he works as a Test and Measurement engineer at u-blox Italia S.p.A.
\end{biography}

\begin{biography}
   [{\includegraphics[width=1in,height=1.25in,clip,keepaspectratio]{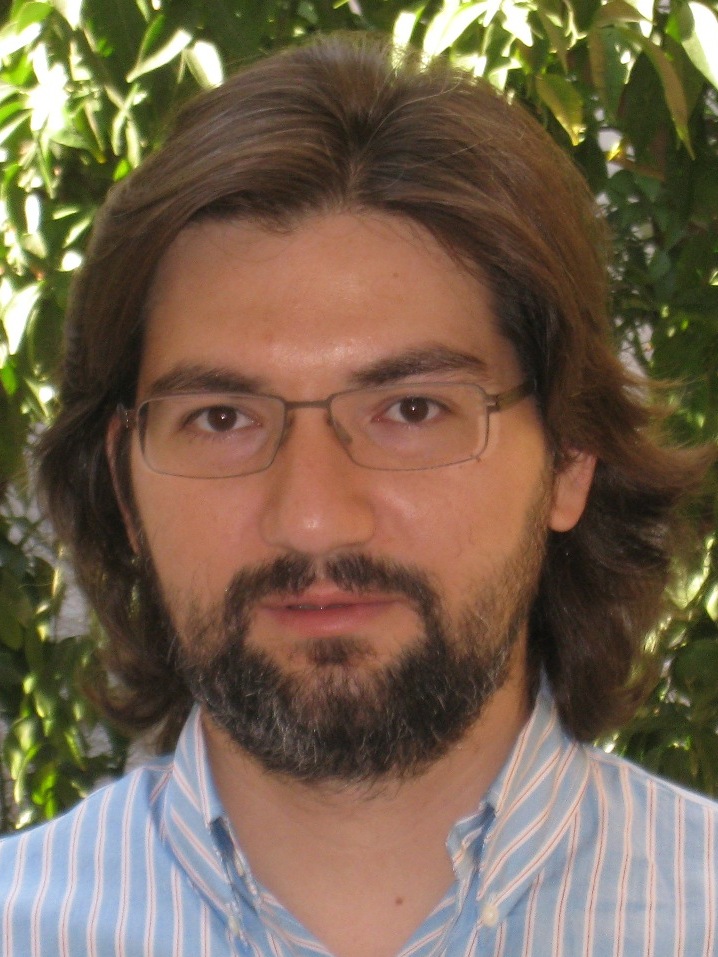}}]{Stavros Toumpis}
Stavros Toumpis received the Diploma in Electrical and Computer Engineering from the National Technical University of Athens, Greece, in 1997, the M.S. degrees in Electrical Engineering and Mathematics from Stanford University, CA, in 1999 and 2002, respectively, and the Ph.D. degree in Electrical Engineering, also from Stanford University, in 2003. He is currently an Assistant Professor at the Department of Informatics of the Athens University of Economics and Business, Athens, Greece. His research is on wireless ad hoc networks, with an emphasis on their capacity and performance evaluation, using tools from network optimization, probability theory, and information theory.
\end{biography}

\begin{biography}
    [{\includegraphics[width=1in,height=1.25in,clip,keepaspectratio]{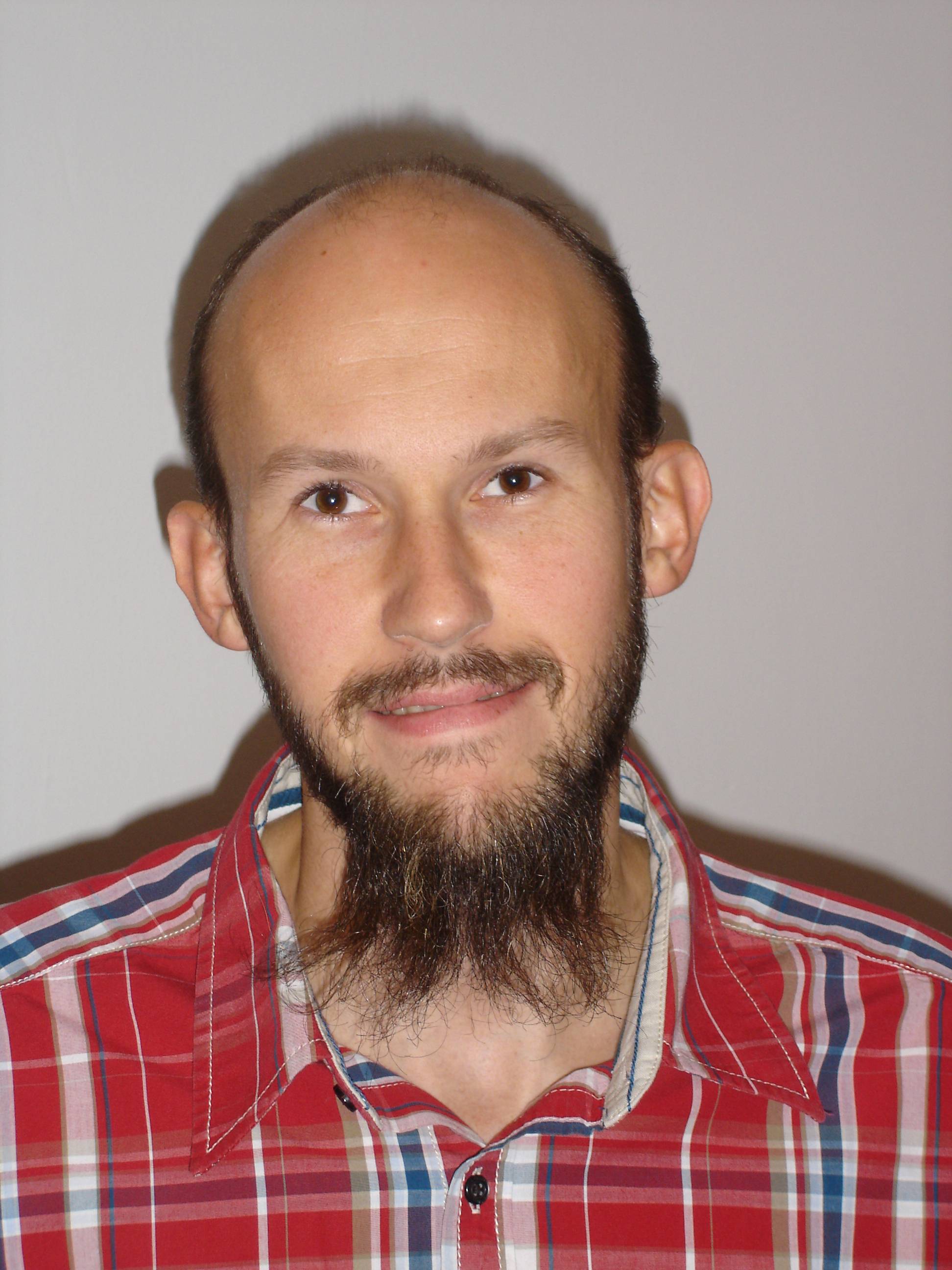}}]{Udo Schilcher}
Udo Schilcher studied applied computing and
mathematics at the University of Klagenfurt,
where he received two Dipl.-Ing. degrees with
distinction (2005, 2006). Since 2005, he has
been research staff member at the Networked
and Embedded Systems institute at the University
of Klagenfurt. His main interests are interference
and node distributions in wireless networks.
His doctoral thesis on inhomogeneous
node distributions and interference correlation in
wireless networks and has been awarded with
a Dr. techn. degree with distinction in 2011. He received a best paper
award from the IEEE Vehicular Technology Society.
\end{biography}

\begin{biography}
    [{\includegraphics[width=1in,height=1.25in,clip,keepaspectratio]{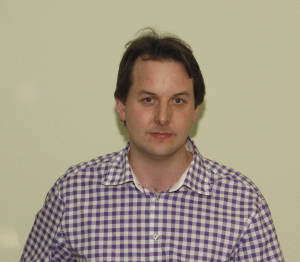}}]{G\"{u}nther Brandner}
G\"{u}nther Brandner studied applied computing
and mathematics at the University of Klagenfurt,
where he received two Dipl.-Ing. degrees
with distinction (2007, 2008). Since 2007 he
has been a research and teaching staff member
and doctoral student at the Networked and
Embedded Systems institute at the University of
Klagenfurt. His main interests are relay selection
methods for cooperative relaying and the
implementation and evaluation of protocols on
hardware platforms.
\end{biography}

\begin{biography}
    [{\includegraphics[width=1in,height=1.25in,clip,keepaspectratio]{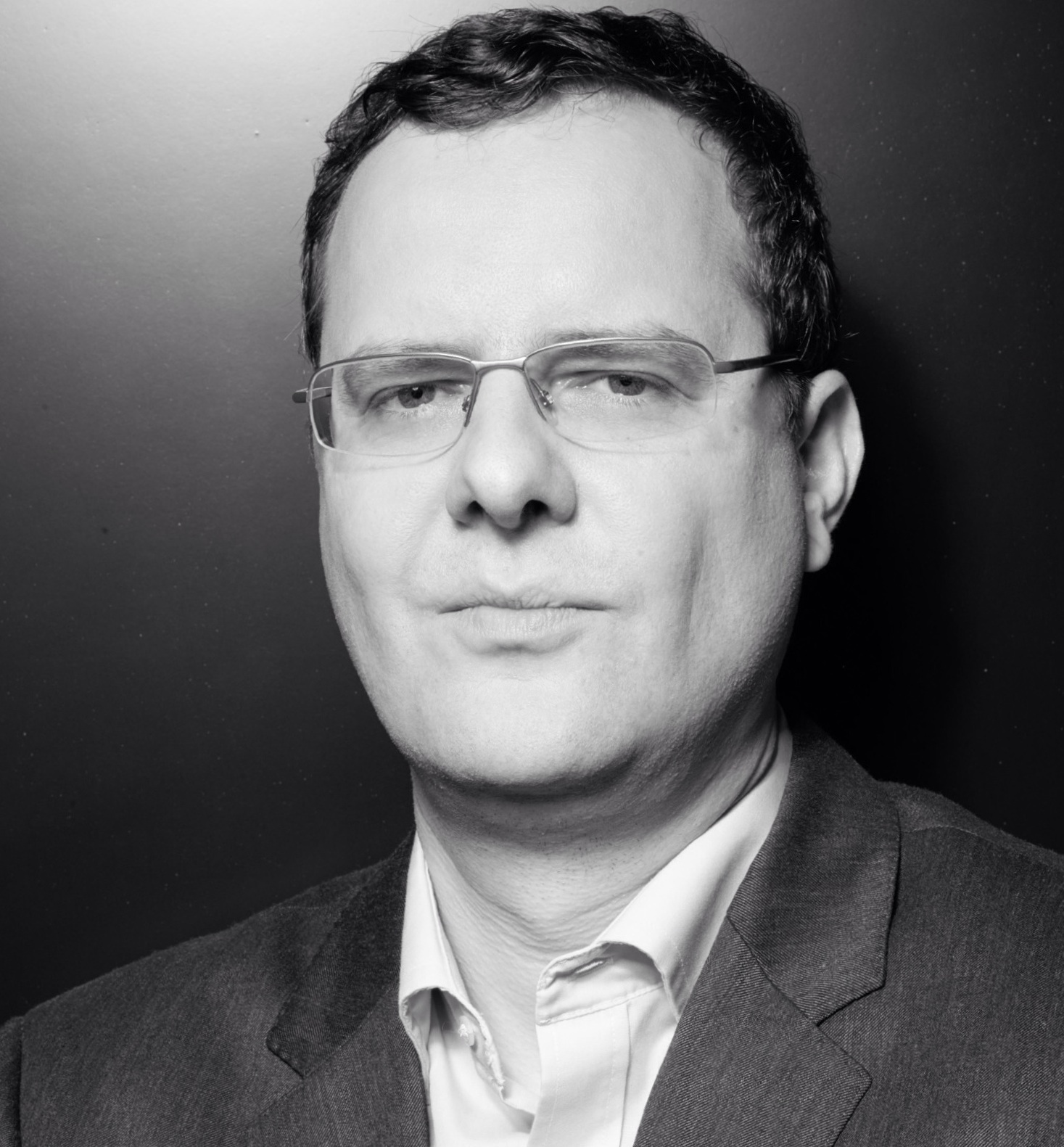}}]{Christian Bettstetter}
Christian Bettstetter (S'98-M'04-SM'09) received the Dipl.-Ing. degree in 1998 and the Dr.-Ing. degree (summa cum laude) in 2004, both in electrical engineering and information technology from the Technische Universit\"{a}t M\"{u}nchen (TUM), Munich, Germany.

He was a Staff Member with the Communications Networks Institute, TUM, until 2003. From 2003 to 2005, he was a Senior Researcher with the DOCOMO Euro-Labs. Since 2005, he has been a Professor and Head of the Institute of Networked and Embedded Systems, University of Klagenfurt, Austria. He is also Scientific Director and Founder of Lakeside Labs, Klagenfurt, a research cluster on self-organizing networked systems. He coauthored the textbook GSM-Architecture, Protocols and Services (Wiley).

Mr. Bettstetter received Best Paper Awards from the IEEE Vehicular Technology Society and the German Information Technology Society (ITG).
\end{biography}

\end{document}

%% file: Figures/dir-vs-single-relay-psucc.tex
\begin{tikzpicture}
    \begin{axis}[
                scale=1.0,
                xlabel={Transmission probability $\txp$},
                ylabel={Success probability $\Psucc$},
                xmin=0, xmax=1, ymin=0, ymax=1,
                xtick={0,0.25,...,1}, ytick={0,0.2,0.4,0.6,0.8,1},
                xticklabels={0,0.25,...,1}, yticklabels={0,0.2,0.4,0.6,0.8,1},
                grid=both,
                legend style={at={(0,0)}, anchor=south west, font=\footnotesize},
                legend cell align=left,
                legend columns=1,
                font=\footnotesize
            ]
            \addplot[color=black, mark=star, mark options={solid, scale=1.2}, mark repeat=1] table[x index =0, y index =2] {Results/single-relay-sel.dat};
            \label{plt:dir-vs-single-sel-lam-0dot2}
            \addplot[color=black, mark=+, mark options={solid, scale=1.2}, mark repeat=1] table[x index =0, y index =6] {Results/single-relay-sel.dat};
            \label{plt:dir-vs-single-sel-lam-0dot6}
            \addplot[color=black, mark=diamond, mark options={solid, scale=1.2}, mark repeat=1] table[x index =0, y index =10] {Results/single-relay-sel.dat};
            \label{plt:dir-vs-single-sel-lam-1dot0}
            \addplot[color=black, mark=o, mark options={solid, scale=1.2}, mark repeat=1] table[x index =0, y index =20] {Results/single-relay-sel.dat};
            \label{plt:dir-vs-single-sel-lam-2dot0}
            \addplot[color=black, dashed, mark=star, mark options={solid, scale=1.2}, mark repeat=1] table[x index =0, y index =2] {Results/direct-psucc-txp.dat};
            \label{plt:dir-vs-single-dir-lam-0dot2}
            \addplot[color=black, dashed, mark=+, mark options={solid, scale=1.2}, mark repeat=1] table[x index =0, y index =6] {Results/direct-psucc-txp.dat};
            \label{plt:dir-vs-single-dir-lam-0dot6}
            \addplot[color=black, dashed, mark=diamond, mark options={solid, scale=1.2}, mark repeat=1] table[x index =0, y index =10] {Results/direct-psucc-txp.dat};
            \label{plt:dir-vs-single-dir-lam-1dot0}
            \addplot[color=black, dashed, mark=o, mark options={solid, scale=1.2}, mark repeat=1] table[x index =0, y index =20] {Results/direct-psucc-txp.dat};
            \label{plt:dir-vs-single-dir-lam-2dot0}

    \end{axis}
\end{tikzpicture}

%% file: Figures/single-relay-psucc.tex
\begin{tikzpicture}
    \begin{axis}[
                scale=1.0,
                xlabel={Transmission probability $\txp$},
                ylabel={Success probability $\Psucc$},
                xmin=0, xmax=1, ymin=0, ymax=1,
                xtick={0,0.25,...,1}, ytick={0,0.2,0.4,0.6,0.8,1},
                xticklabels={0,0.25,...,1}, yticklabels={0,0.2,0.4,0.6,0.8,1},
                grid=both,
                legend style={at={(0,0)}, anchor=south west, font=\footnotesize},
                legend cell align=left,
                legend columns=1,
                font=\footnotesize
            ]
            \addplot[color=black, mark=star, mark options={solid, scale=1.2}, mark repeat=1] table[x index =0, y index =2] {Results/single-relay-sel.dat};
            \label{plt:single-sel-lam-0dot2}
            \addplot[color=black, mark=+, mark options={solid, scale=1.2}, mark repeat=1] table[x index =0, y index =6] {Results/single-relay-sel.dat};
            \label{plt:single-sel-lam-0dot6}
            \addplot[color=black, mark=diamond, mark options={solid, scale=1.2}, mark repeat=1] table[x index =0, y index =10] {Results/single-relay-sel.dat};
            \label{plt:single-sel-lam-1dot0}
            \addplot[color=black, mark=o, mark options={solid, scale=1.2}, mark repeat=1] table[x index =0, y index =20] {Results/single-relay-sel.dat};
            \label{plt:single-sel-lam-2dot0}
            \addplot[color=black, dashed, mark=star, mark options={solid, scale=1.2}, mark repeat=1] table[x index =0, y index =2] {Results/single-relay-mrc.dat};
            \label{plt:single-mrc-lam-0dot2}
            \addplot[color=black, dashed, mark=+, mark options={solid, scale=1.2}, mark repeat=1] table[x index =0, y index =6] {Results/single-relay-mrc.dat};
            \label{plt:single-mrc-lam-0dot6}
            \addplot[color=black, dashed, mark=diamond, mark options={solid, scale=1.2}, mark repeat=1] table[x index =0, y index =10] {Results/single-relay-mrc.dat};
            \label{plt:single-mrc-lam-1dot0}
            \addplot[color=black, dashed, mark=o, mark options={solid, scale=1.2}, mark repeat=1] table[x index =0, y index =20] {Results/single-relay-mrc.dat};
            \label{plt:single-mrc-lam-2dot0}

            \node[anchor=west] at (axis cs: 0.25,0.07) {$\dens = 2$};
            \node[anchor=west] at (axis cs: 0.5,0.15) {$\dens = 1$};
            \node[anchor=west] at (axis cs: 0.5,0.55) {$\dens = 0.6$};
            \node[anchor=west] at (axis cs: 0.5,0.7) {$\dens = 0.2$};
    \end{axis}
\end{tikzpicture}

%% file: Figures/sel-vs-mrc.tex
\begin{tikzpicture}
    \begin{axis}[
                scale=1.0,
                xlabel={Location of the relay cluster $\cluster$},
                ylabel={Success probability $\Psucc$},
                xmin=0, xmax=1, ymin=0, ymax=1,
                xtick={0,0.25,...,1}, ytick={0,0.2,0.4,0.6,0.8,1},
                xticklabels={0,0.25,...,1}, yticklabels={0,0.2,0.4,0.6,0.8,1},
                grid=both,
                legend style={at={(0,0)}, anchor=south west, font=\footnotesize},
                legend cell align=left,
                legend columns=1,
                font=\footnotesize
            ]
            \addplot[color=black] table[x index =0, y index =1] {Results/sel-th-0dot1-lambda-0dot5-p-1.dat}; \label{plt:sel-vs-mrc-sel-good-N-1}
            \addplot[dashed, color=black] table[x index =0, y index =3] {Results/sel-th-0dot1-lambda-0dot5-p-1.dat}; \label{plt:sel-vs-mrc-sel-good-N-3}
            \addplot[dotted, color=black] table[x index =0, y index =5] {Results/sel-th-0dot1-lambda-0dot5-p-1.dat}; \label{plt:sel-vs-mrc-sel-good-N-5}
            \addplot[color=black, only marks, mark=star, mark options={scale=.95}, mark repeat=2] table[x index =0, y index =1] {Results/mrc-th-0dot1-lambda-0dot5-p-1.dat}; \label{plt:sel-vs-mrc-mrc-good-N-1}
            \addplot[color=black, only marks, mark=o, mark options={scale=.75}, mark repeat=2] table[x index =0, y index =3] {Results/mrc-th-0dot1-lambda-0dot5-p-1.dat}; \label{plt:sel-vs-mrc-mrc-good-N-3}
            \addplot[color=black, only marks, mark=+, mark options={scale=.95}, mark repeat=2] table[x index =0, y index =5] {Results/mrc-th-0dot1-lambda-0dot5-p-1.dat}; \label{plt:sel-vs-mrc-mrc-good-N-5}
            \addplot[thick, color=black] table[x index =0, y index =1] {Results/sel-th-1-lambda-1-p-1.dat};
            \addplot[thick, dashed, color=black] table[x index =0, y index =3] {Results/sel-th-1-lambda-1-p-1.dat};
            \addplot[thick, dotted, color=black] table[x index =0, y index =5] {Results/sel-th-1-lambda-1-p-1.dat};
            \addplot[thick, color=black, only marks, mark=star, mark options={scale=.95}, mark repeat=2] table[x index =0, y index =1] {Results/mrc-th-1-lambda-1-p-1.dat};
            \addplot[thick, color=black, only marks, mark=o, mark options={scale=.75}, mark repeat=2] table[x index =0, y index =3] {Results/mrc-th-1-lambda-1-p-1.dat};
            \addplot[thick, color=black, only marks, mark=+, mark options={scale=.95}, mark repeat=2] table[x index =0, y index =5] {Results/mrc-th-1-lambda-1-p-1.dat};

            \node[anchor=west] at (axis cs: 0.015,0.25) {harsh scenario};
            \node at (axis cs: 0.5,0.05) {$N=1$};
            \node at (axis cs: 0.5,0.2) {$N=3$};
            \node at (axis cs: 0.5,0.35) {$N=5$};
            \node[anchor=west] at (axis cs: 0.015,0.9) {good scenario};
            \node at (axis cs: 0.5,0.7) {$N=1$};
            \node at (axis cs: 0.5,0.8) {$N=3$};
            \node at (axis cs: 0.5,0.91) {$N=5$};

    \end{axis}
\end{tikzpicture}

%% file: Figures/random-pos.tex
\begin{tikzpicture}
    \begin{axis}[
                scale=1.0,
                xlabel={Square side length $\edge$},
                ylabel={Success probability $\Psucc$},
                xmin=0, xmax=1, ymin=0, ymax=1,
                xtick={0,0.2,...,1}, ytick={0,0.2,0.4,0.6,0.8,1},
                xticklabels={0,0.2,0.4,0.6,0.8,1}, yticklabels={0,0.2,0.4,0.6,0.8,1},
                grid=both,
                legend style={at={(0,0)}, anchor=south west, font=\footnotesize},
                legend cell align=left,
                legend columns=1,
                font=\footnotesize
            ]
            \addplot[color=black] table[x index =0, y index =1] {Results/random-mrc-0dot5-th-0dot1-lambda-0dot5-p-1.dat}; \label{plt:random-mrc-good-N-1}
            \addplot[color=black, dashdotted] table[x index =0, y index =2] {Results/random-mrc-0dot5-th-0dot1-lambda-0dot5-p-1.dat}; \label{plt:random-mrc-good-N-2}
            \addplot[densely dotted, color=black] table[x index =0, y index =3] {Results/random-mrc-0dot5-th-0dot1-lambda-0dot5-p-1.dat}; \label{plt:random-mrc-good-N-3}
            \addplot[dashed, color=black] table[x index =0, y index =4] {Results/random-mrc-0dot5-th-0dot1-lambda-0dot5-p-1.dat}; \label{plt:random-mrc-good-N-4}
            \addplot[dotted, color=black] table[x index =0, y index =5] {Results/random-mrc-0dot5-th-0dot1-lambda-0dot5-p-1.dat}; \label{plt:random-mrc-good-N-5}
            \addplot[color=black, very thick] table[x index =0, y index =1] {Results/random-mrc-0dot5-th-1-lambda-1-p-1.dat};
            \addplot[color=black, very thick, dashdotted] table[x index =0, y index =2] {Results/random-mrc-0dot5-th-1-lambda-1-p-1.dat};
            \addplot[color=black, very thick, densely dotted] table[x index =0, y index =3] {Results/random-mrc-0dot5-th-1-lambda-1-p-1.dat};
            \addplot[color=black, very thick, dashed] table[x index =0, y index =4] {Results/random-mrc-0dot5-th-1-lambda-1-p-1.dat};
            \addplot[color=black, very thick, dotted] table[x index =0, y index =5] {Results/random-mrc-0dot5-th-1-lambda-1-p-1.dat};

            \node[anchor=west] at (axis cs: 0.65,0.3) {harsh scenario};
            \node[anchor=west] at (axis cs: 0.65,0.95) {good scenario};
    \end{axis}
\end{tikzpicture}

%% file: Figures/succ-vs-th.tex
\begin{tikzpicture}
    \begin{semilogxaxis}[
                scale=1.0,
                xlabel={Success threshold $\thr$},
                ylabel={Success probability $\Psucc$},
                xmin=0.1, xmax=2, ymin=0, ymax=1,
                xtick={0.1,0.2,0.4,0.6,0.8,1,1.2,1.4,1.6,1.8,2}, ytick={0,0.2,0.4,0.6,0.8,1},
                xticklabels={0.1,0.2,0.4,0.6,0.8,1,,1.4,,,2}, yticklabels={0,0.2,0.4,0.6,0.8,1},
                grid=both,
                legend style={at={(0,0)}, anchor=south west, font=\footnotesize},
                legend cell align=left,
                legend columns=1,
                font=\footnotesize
             ]
             \addplot[dashed, color=black, mark=square, mark options={scale=1, solid}, mark repeat=2] table[x index =0, y index =2] {Results/psucc-vs-theta-mrc-lambda-0dot1-p-0dot5.dat}; \label{plt:succ-vs-th-lambda-0dot1-p-0dot5-N-1};
             \addplot[dashed, color=black, mark=diamond, mark options={scale=1.0, solid}, mark repeat=2] table[x index =0, y index =4] {Results/psucc-vs-theta-mrc-lambda-0dot1-p-0dot5.dat}; \label{plt:succ-vs-th-lambda-0dot1-p-0dot5-N-3};
             \addplot[dashed, color=black, mark=triangle, mark options={scale=1.0, solid}, mark repeat=2] table[x index =0, y index =6] {Results/psucc-vs-theta-mrc-lambda-0dot1-p-0dot5.dat}; \label{plt:succ-vs-th-lambda-0dot1-p-0dot5-N-5};
             \addplot[color=black, mark=square, mark options={scale=1}, mark repeat=2] table[x index =0, y index =2] {Results/psucc-vs-theta-mrc-lambda-1-p-1.dat}; \label{plt:succ-vs-th-lambda-1-p-1-N-1};
             \addplot[color=black, mark=diamond, mark options={scale=1.0}, mark repeat=2] table[x index =0, y index =4] {Results/psucc-vs-theta-mrc-lambda-1-p-1.dat}; \label{plt:succ-vs-th-lambda-1-p-1-N-3};
             \addplot[color=black, mark=triangle, mark options={scale=1.0}, mark repeat=2] table[x index =0, y index =6] {Results/psucc-vs-theta-mrc-lambda-1-p-1.dat}; \label{plt:succ-vs-th-lambda-1-p-1-N-5};
             \addplot[dotted, color=black, mark=square, mark options={scale=1, solid}, mark repeat=2] table[x index =0, y index =2] {Results/psucc-vs-theta-mrc-lambda-0dot5-p-0dot75.dat}; \label{plt:succ-vs-th-lambda-0dot5-p-0dot75-N-1};
             \addplot[dotted, color=black, mark=diamond, mark options={scale=1.0, solid}, mark repeat=2] table[x index =0, y index =4] {Results/psucc-vs-theta-mrc-lambda-0dot5-p-0dot75.dat}; \label{plt:succ-vs-th-lambda-0dot5-p-0dot75-N-3};
             \addplot[dotted, color=black, mark=triangle, mark options={scale=1.0, solid}, mark repeat=2] table[x index =0, y index =6] {Results/psucc-vs-theta-mrc-lambda-0dot5-p-0dot75.dat}; \label{plt:succ-vs-th-lambda-0dot5-p-0dot75-N-5};
             \addplot[only marks, color=black, mark=*, mark options={scale=.45}, mark repeat=1] table[x index =2, y index =3] {Results/sim-relay-mrc-lambda-0dot1-p-0dot5-N-1.dat}; \label{plt:sim-relay-mrc-lambda-0dot1-p-0dot5-N-1};
             \addplot[only marks, color=black, mark=*, mark options={scale=.45}, mark repeat=1] table[x index =2, y index =3] {Results/sim-relay-mrc-lambda-0dot1-p-0dot5-N-3.dat}; \label{plt:sim-relay-mrc-lambda-0dot1-p-0dot5-N-3};
             \addplot[only marks, color=black, mark=*, mark options={scale=.45}, mark repeat=1] table[x index =2, y index =3] {Results/sim-relay-mrc-lambda-0dot1-p-0dot5-N-5.dat}; \label{plt:sim-relay-mrc-lambda-0dot1-p-0dot5-N-5};
             \addplot[only marks, color=black, mark=*, mark options={scale=.45}, mark repeat=1] table[x index =2, y index =3] {Results/sim-relay-mrc-lambda-0dot5-p-0dot75-N-1.dat}; \label{plt:sim-relay-mrc-lambda-0dot5-p-0dot75-N-1};
             \addplot[only marks, color=black, mark=*, mark options={scale=.45}, mark repeat=1] table[x index =2, y index =3] {Results/sim-relay-mrc-lambda-0dot5-p-0dot75-N-3.dat}; \label{plt:sim-relay-mrc-lambda-0dot5-p-0dot75-N-3};
             \addplot[only marks, color=black, mark=*, mark options={scale=.45}, mark repeat=1] table[x index =2, y index =3] {Results/sim-relay-mrc-lambda-0dot5-p-0dot75-N-5.dat}; \label{plt:sim-relay-mrc-lambda-0dot5-p-0dot75-N-5};
             \addplot[only marks, color=black, mark=*, mark options={scale=.45}, mark repeat=1] table[x index =2, y index =3] {Results/sim-relay-mrc-lambda-1-p-1-N-1.dat}; \label{plt:sim-relay-mrc-lambda-1-p-1-N-1};
             \addplot[only marks, color=black, mark=*, mark options={scale=.45}, mark repeat=1] table[x index =2, y index =3] {Results/sim-relay-mrc-lambda-1-p-1-N-3.dat}; \label{plt:sim-relay-mrc-lambda-1-p-1-N-3};
             \addplot[only marks, color=black, mark=*, mark options={scale=.45}, mark repeat=1] table[x index =2, y index =3] {Results/sim-relay-mrc-lambda-1-p-1-N-5.dat}; \label{plt:sim-relay-mrc-lambda-1-p-1-N-5};
    \end{semilogxaxis}
\end{tikzpicture}

%% file: Figures/psucc-times-rate-vs-th.tex
\begin{tikzpicture}
    \begin{axis}[
                scale=1.0,
                xlabel={Success threshold $\thr$},
                ylabel={$\Psucc \; \ln (1 + \thr)$},
                xmin=0, xmax=10, ymin=0, ymax=2,
                grid=both,
                legend style={at={(0,0)}, anchor=south west, font=\footnotesize},
                legend cell align=left,
                legend columns=1,
                font=\footnotesize
             ]
             \addplot[dashed, color=black, mark=+, mark options={scale=.85, solid}, mark repeat=4] table[x index =0, y index =1] {Results/relay-mrc-const-lambda-0dot75-p-0dot5.dat}; \label{plt:relay-mrc-const-lambda-0dot75-p-0dot5-N-1};
             \addplot[dashed, color=black, mark=*, mark options={scale=.65, solid}, mark repeat=4] table[x index =0, y index =2] {Results/relay-mrc-const-lambda-0dot75-p-0dot5.dat}; \label{plt:relay-mrc-const-lambda-0dot75-p-0dot5-N-3};
             \addplot[dashed, color=black, mark=diamond, mark options={scale=.85, solid}, mark repeat=4] table[x index =0, y index =3] {Results/relay-mrc-const-lambda-0dot75-p-0dot5.dat}; \label{plt:relay-mrc-const-lambda-0dot75-p-0dot5-N-5};
             \addplot[color=black, mark=+, mark options={scale=.85, solid}, mark repeat=4] table[x index =0, y index =1] {Results/relay-mrc-const-lambda-1-p-1.dat}; \label{plt:relay-mrc-const-lambda-1-p-1-N-1};
             \addplot[color=black, mark=*, mark options={scale=.65, solid}, mark repeat=4] table[x index =0, y index =2] {Results/relay-mrc-const-lambda-1-p-1.dat}; \label{plt:relay-mrc-const-lambda-1-p-1-N-3};
             \addplot[color=black, mark=diamond, mark options={scale=.85, solid}, mark repeat=4] table[x index =0, y index =3] {Results/relay-mrc-const-lambda-1-p-1.dat}; \label{plt:relay-mrc-const-lambda-1-p-1-N-5};
             \addplot[dotted, color=black, mark=+, mark options={scale=.85, solid}, mark repeat=4] table[x index =0, y index =1] {Results/relay-mrc-const-lambda-0dot25-p-0dot75.dat}; \label{plt:relay-mrc-const-lambda-0dot25-p-0dot75-N-1};
             \addplot[dotted, color=black, mark=*, mark options={scale=.65, solid}, mark repeat=4] table[x index =0, y index =2] {Results/relay-mrc-const-lambda-0dot25-p-0dot75.dat}; \label{plt:relay-mrc-const-lambda-0dot25-p-0dot75-N-3};
             \addplot[dotted, color=black, mark=diamond, mark options={scale=.85, solid}, mark repeat=4] table[x index =0, y index =3] {Results/relay-mrc-const-lambda-0dot25-p-0dot75.dat}; \label{plt:relay-mrc-const-lambda-0dot25-p-0dot75-N-5};
             \addplot[dashdotted, color=black, mark=+, mark options={scale=.85, solid}, mark repeat=4] table[x index =0, y index =1] {Results/relay-mrc-const-lambda-0dot1-p-0dot5.dat}; \label{plt:relay-mrc-const-lambda-0dot1-p-0dot5-N-1};
             \addplot[dashdotted, color=black, mark=*, mark options={scale=.65, solid}, mark repeat=4] table[x index =0, y index =2] {Results/relay-mrc-const-lambda-0dot1-p-0dot5.dat}; \label{plt:relay-mrc-const-lambda-0dot1-p-0dot5-N-3};
             \addplot[dashdotted, color=black, mark=diamond, mark options={scale=.85, solid}, mark repeat=4] table[x index =0, y index =3] {Results/relay-mrc-const-lambda-0dot1-p-0dot5.dat}; \label{plt:relay-mrc-const-lambda-0dot1-p-0dot5-N-5};
    \end{axis}
\end{tikzpicture}

%% file: Figures/renewal-ind-vs-dep.tex
\begin{tikzpicture}
    \begin{axis}[
        scale=1.0,
        xlabel={Transmission attempt $t$},
        ylabel={cdf $P[T \leq t]$},
        xmin=0,xmax=6, ymin=0,ymax=1,
        xtick={1,...,5}, ytick={0,0.25,...,1},
        xticklabels={1,...,5}, yticklabels={0,0.25,...,1},
        grid=both,
        legend style={at={(1,0)}, anchor=south east, font=\footnotesize},
        legend cell align=left,
        legend columns=1,
        font=\footnotesize
    ]
    \addplot[color=black] table[x index =0, y index =4] {Results/renewal-cdf-mrc-th-0dot1-lambda-0dot5-p-1.dat}; \label{plt:renewal-mrc-dep-good};
    \addplot[color=black, very thick] table[x index =0, y index =4] {Results/renewal-cdf-mrc-th-1-lambda-1-p-1.dat};
    \addplot[color=black, dashed] table[x index =0, y index =4] {Results/renewal-cdf-mrc-ind-th-0dot1-lambda-0dot5-p-1.dat}; \label{plt:renewal-mrc-ind-good};
    \addplot[color=black, dashed, very thick] table[x index =0, y index =4] {Results/renewal-cdf-mrc-ind-th-1-lambda-1-p-1.dat};
    \addplot[color=black, dotted, very thick] table[x index =0, y index =2] {Results/sim-cdf-corr-mrc-lambda-1-p-1-th-1-speed-0dot1-N-3.dat}; \label{plt:sim-renewal-mrc-speed-0dot1-bad};
    \addplot[color=black, dotted, very thick] table[x index =0, y index =2] {Results/sim-cdf-corr-mrc-lambda-1-p-1-th-1-speed-0dot2-N-3.dat}; \label{plt:sim-renewal-mrc-speed-0dot2-bad};
    \addplot[color=black, dotted, very thick] table[x index =0, y index =2] {Results/sim-cdf-corr-mrc-lambda-1-p-1-th-1-speed-0dot5-N-3.dat}; \label{plt:sim-renewal-mrc-speed-0dot5-bad};
    \addplot[color=black, dotted, very thick] table[x index =0, y index =2] {Results/sim-cdf-corr-mrc-lambda-1-p-1-th-1-speed-0dot75-N-3.dat}; \label{plt:sim-renewal-mrc-speed-0dot75-bad};
    \addplot[color=black, dotted, very thick] table[x index =0, y index =2] {Results/sim-cdf-corr-mrc-lambda-1-p-1-th-1-speed-1-N-3.dat}; \label{plt:sim-renewal-mrc-speed-1-bad};
    \addplot[color=black, dotted, very thick] table[x index =0, y index =2] {Results/sim-cdf-corr-mrc-lambda-1-p-1-th-1-speed-2-N-3.dat}; \label{plt:sim-renewal-mrc-speed-2-bad};
    \addplot[color=black, dotted] table[x index =0, y index =2] {Results/sim-cdf-corr-mrc-lambda-0dot5-p-1-th-0dot1-speed-0dot1-N-3.dat}; \label{plt:sim-renewal-mrc-speed-0dot1-good};
    \addplot[color=black, dotted] table[x index =0, y index =2] {Results/sim-cdf-corr-mrc-lambda-0dot5-p-1-th-0dot1-speed-0dot2-N-3.dat}; \label{plt:sim-renewal-mrc-speed-0dot2-good};
    \addplot[color=black, dotted] table[x index =0, y index =2] {Results/sim-cdf-corr-mrc-lambda-0dot5-p-1-th-0dot1-speed-0dot5-N-3.dat}; \label{plt:sim-renewal-mrc-speed-0dot5-good};
    \addplot[color=black, dotted] table[x index =0, y index =2] {Results/sim-cdf-corr-mrc-lambda-0dot5-p-1-th-0dot1-speed-0dot75-N-3.dat}; \label{plt:sim-renewal-mrc-speed-0dot75-good};
    \addplot[color=black, dotted] table[x index =0, y index =2] {Results/sim-cdf-corr-mrc-lambda-0dot5-p-1-th-0dot1-speed-1-N-3.dat}; \label{plt:sim-renewal-mrc-speed-1-good};
    \addplot[color=black, dotted] table[x index =0, y index =2] {Results/sim-cdf-corr-mrc-lambda-0dot5-p-1-th-0dot1-speed-2-N-3.dat}; \label{plt:sim-renewal-mrc-speed-2-good};

    \node[anchor=west] at (axis cs: 2.5,0.25) {harsh scenario};
    \node[anchor=west] at (axis cs: 2.5,0.85) {good scenario};

    \draw[thick, ->] (axis cs: 2.9,0.025) -- (axis cs: 2.9,0.15) ;
    \node at (axis cs: 4.4,0.1) {$\speed = 0.1, 0.2, 0.5, 0.75, 1, 2$};

    \end{axis}
\end{tikzpicture}

%% file: Figures/renewal-num-relays.tex
\begin{tikzpicture}
    \begin{axis}[
        scale=1.0,
        xlabel={Transmission attempt $t$},
        ylabel={cdf $P[T \leq t]$},
        xmin=0,xmax=6, ymin=0,ymax=1,
        xtick={1,...,5}, ytick={0,0.25,...,1},
        xticklabels={1,...,5}, yticklabels={0,0.25,...,1},
        grid=both,
        legend style={at={(1,0)}, anchor=south east, font=\footnotesize},
        legend cell align=left,
        legend columns=1,
        font=\footnotesize
    ]
    \addplot[color=black] table[x index =0, y index =1] {Results/renewal-cdf-mrc-th-0dot1-lambda-0dot5-p-1.dat}; \label{plt:renewal-mrc-good-N-0};
    \addplot[color=black, dashdotted] table[x index =0, y index =2] {Results/renewal-cdf-mrc-th-0dot1-lambda-0dot5-p-1.dat}; \label{plt:renewal-mrc-good-N-1};
    \addplot[color=black, dashed] table[x index =0, y index =3] {Results/renewal-cdf-mrc-th-0dot1-lambda-0dot5-p-1.dat}; \label{plt:renewal-mrc-good-N-2};
    \addplot[color=black, dotted] table[x index =0, y index =4] {Results/renewal-cdf-mrc-th-0dot1-lambda-0dot5-p-1.dat}; \label{plt:renewal-mrc-good-N-3};
    \addplot[color=black, very thick] table[x index =0, y index =1] {Results/renewal-cdf-mrc-th-1-lambda-1-p-1.dat};
    \addplot[color=black, dashdotted, very thick] table[x index =0, y index =2] {Results/renewal-cdf-mrc-th-1-lambda-1-p-1.dat};
    \addplot[color=black, dashed, very thick] table[x index =0, y index =3] {Results/renewal-cdf-mrc-th-1-lambda-1-p-1.dat};
    \addplot[color=black, dotted, very thick] table[x index =0, y index =4] {Results/renewal-cdf-mrc-th-1-lambda-1-p-1.dat};

    \node[anchor=west] at (axis cs: 2.5,0.45) {harsh scenario};
    \node[anchor=west] at (axis cs: 2.5,0.97) {good scenario};

    \end{axis}
\end{tikzpicture}

%% file: Figures/renewal-throughput-th.tex
\begin{tikzpicture}
    \begin{axis}[
                scale=1.0,
                xlabel={Success threshold $\thr$},
                ylabel={Throughput $\Lambda$},
                xmin=0, xmax=2, ymin=0, ymax=1,
                xtick={0, 0.2,0.4,0.6,0.8,1,1.2,1.4,1.6,1.8,2}, ytick={0,0.2,0.4,0.6,0.8,1},
                xticklabels={0.0, 0.2,0.4,0.6,0.8,1,1.2,1.4,1.6,1.8,2}, yticklabels={0,0.2,0.4,0.6,0.8,1},
                grid=both,
                legend style={at={(0,0)}, anchor=south west, font=\footnotesize},
                legend cell align=left,
                legend columns=1,
                font=\footnotesize
             ]
             \addplot[color=black] table[x index =0, y index =1] {Results/throughput-vs-th-const-ppp-mrc-lambda-0dot25-p-1.dat}; \label{plt:throughput-vs-th-const-ppp-mrc-lambda-0dot25-p-1-N-1};
             \addplot[color=black, dashed] table[x index =0, y index =1] {Results/throughput-vs-th-ind-ppp-mrc-lambda-0dot25-p-1.dat}; \label{plt:throughput-vs-th-ind-ppp-mrc-lambda-0dot25-p-1-N-1};
             \addplot[very thick, color=black] table[x index =0, y index =1] {Results/throughput-vs-th-const-ppp-mrc-lambda-1-p-1.dat}; \label{plt:throughput-vs-th-const-ppp-mrc-lambda-1-p-1-N-1};
             \addplot[very thick, color=black, dashed] table[x index =0, y index =1] {Results/throughput-vs-th-ind-ppp-mrc-lambda-1-p-1.dat}; \label{plt:throughput-vs-th-ind-ppp-mrc-lambda-1-p-1-N-1};
             \addplot[color=black, only marks, mark=*, mark options={scale=.4}, mark repeat=1] table[x index =2, y index =3] {Results/sim-throughput-vs-th-const-ppp-mrc-lambda-0dot25-p-1-N-1.dat}; \label{plt:sim-throughput-vs-th-const-ppp-mrc-lambda-0dot25-p-1-N-1};
             \addplot[color=black, only marks, mark=*, mark options={scale=.4}, mark repeat=1] table[x index =2, y index =3] {Results/sim-throughput-vs-th-ind-ppp-mrc-lambda-0dot25-p-1-N-1.dat}; \label{plt:sim-throughput-vs-th-ind-ppp-mrc-lambda-0dot25-p-1-N-1};
             \addplot[color=black, only marks, mark=*, mark options={scale=.4}, mark repeat=1] table[x index =2, y index =3] {Results/sim-throughput-vs-th-const-ppp-mrc-lambda-1-p-1-N-1.dat}; \label{plt:sim-throughput-vs-th-const-ppp-mrc-lambda-1-p-1-N-1};
             \addplot[color=black, only marks, mark=*, mark options={scale=.4}, mark repeat=1] table[x index =2, y index =3] {Results/sim-throughput-vs-th-ind-ppp-mrc-lambda-1-p-1-N-1.dat}; \label{plt:sim-throughput-vs-th-ind-ppp-mrc-lambda-1-p-1-N-1};
             \addplot[thick, color=black, dotted] table[x index =2, y index =3] {Results/sim-throughput-vs-th-corr-ppp-mrc-lambda-1-p-1-N-1-speed-0dot1.dat}; \label{plt:sim-throughput-vs-th-corr-ppp-mrc-lambda-1-p-1-N-1-speed-0dot1};
             \addplot[thick, color=black, dotted] table[x index =2, y index =3] {Results/sim-throughput-vs-th-corr-ppp-mrc-lambda-1-p-1-N-1-speed-0dot2.dat}; \label{plt:sim-throughput-vs-th-corr-ppp-mrc-lambda-1-p-1-N-1-speed-0dot2};
             \addplot[thick, color=black, dotted] table[x index =2, y index =3] {Results/sim-throughput-vs-th-corr-ppp-mrc-lambda-1-p-1-N-1-speed-0dot5.dat}; \label{plt:sim-throughput-vs-th-corr-ppp-mrc-lambda-1-p-1-N-1-speed-0dot5};
             \addplot[thick, color=black, dotted] table[x index =2, y index =3] {Results/sim-throughput-vs-th-corr-ppp-mrc-lambda-1-p-1-N-1-speed-0dot75.dat}; \label{plt:sim-throughput-vs-th-corr-ppp-mrc-lambda-1-p-1-N-1-speed-0dot75};
             \addplot[thick, color=black, dotted] table[x index =2, y index =3] {Results/sim-throughput-vs-th-corr-ppp-mrc-lambda-1-p-1-N-1-speed-1.dat}; \label{plt:sim-throughput-vs-th-corr-ppp-mrc-lambda-1-p-1-N-1-speed-1};
             \addplot[thick, color=black, dotted] table[x index =2, y index =3] {Results/sim-throughput-vs-th-corr-ppp-mrc-lambda-1-p-1-N-1-speed-2.dat}; \label{plt:sim-throughput-vs-th-corr-ppp-mrc-lambda-1-p-1-N-1-speed-2};
             \addplot[color=black, dotted] table[x index =2, y index =3] {Results/sim-throughput-vs-th-corr-ppp-mrc-lambda-0dot25-p-1-N-1-speed-0dot1.dat}; \label{plt:sim-throughput-vs-th-corr-ppp-mrc-lambda-0dot25-p-1-N-1-speed-0dot1};
             \addplot[color=black, dotted] table[x index =2, y index =3] {Results/sim-throughput-vs-th-corr-ppp-mrc-lambda-0dot25-p-1-N-1-speed-0dot2.dat}; \label{plt:sim-throughput-vs-th-corr-ppp-mrc-lambda-0dot25-p-1-N-1-speed-0dot2};
             \addplot[color=black, dotted] table[x index =2, y index =3] {Results/sim-throughput-vs-th-corr-ppp-mrc-lambda-0dot25-p-1-N-1-speed-0dot75.dat}; \label{plt:sim-throughput-vs-th-corr-ppp-mrc-lambda-0dot25-p-1-N-1-speed-0dot75};
             \addplot[color=black, dotted] table[x index =2, y index =3] {Results/sim-throughput-vs-th-corr-ppp-mrc-lambda-0dot25-p-1-N-1-speed-1.dat}; \label{plt:sim-throughput-vs-th-corr-ppp-mrc-lambda-0dot25-p-1-N-1-speed-1};
             \addplot[color=black, dotted] table[x index =2, y index =3] {Results/sim-throughput-vs-th-corr-ppp-mrc-lambda-0dot25-p-1-N-1-speed-2.dat}; \label{plt:sim-throughput-vs-th-corr-ppp-mrc-lambda-0dot25-p-1-N-1-speed-2};

             \node at (axis cs: 1.8,0.65) {$\dens$\,=\,0.25};
             \node at (axis cs: 1.8,0.15) {$\dens$\,=\,1};

             \draw[thick, ->] (axis cs: 0.95,0.8) -- (axis cs: 0.95,0.9) ;
             \node at (axis cs: 1.45,0.85) {$\speed = 0.1, 0.2, 0.5, 0.75, 1, 2$};

    \end{axis}
\end{tikzpicture}

%% file: Figures/renewal-throughput-doppler.tex
\begin{tikzpicture}
    \begin{semilogyaxis}[
                scale=1.0,
                xlabel={Density $\dens$},
                ylabel={Throughput $\Lambda$},
                xmin=0, xmax=2, ymin=1e-2, ymax=1,
                xtick={0,0.2,0.4,0.6,0.8,1,1.2,1.4,1.6,1.8,2}, ytick={1e-2,1e-1,1},
                xticklabels={0,0.2,0.4,0.6,0.8,1,1.2,1.4,1.6,1.8,2}, yticklabels={1e-2,1e-1,1},
                grid=both,
                legend style={at={(0,0)}, anchor=south west, font=\footnotesize},
                legend cell align=left,
                legend columns=1,
                font=\footnotesize
             ]
            \addplot[dashed, color=black, mark=+, mark options={scale=.8, solid}, mark repeat=1] table[x index =0, y index =4] {Results/sim-throughput-vs-lam-corr-fad-const-ppp-mrc-th-0dot25-p-1-N-1-fd-0dot01.dat}; \label{plt:sim-throughput-vs-lam-corr-fad-const-ppp-mrc-th-0dot25-p-1-N-1-fd-0dot01};
            \addplot[dashed, color=black, mark=o, mark options={scale=.8, solid}, mark repeat=1] table[x index =0, y index =4] {Results/sim-throughput-vs-lam-corr-fad-const-ppp-mrc-th-0dot25-p-1-N-1-fd-0dot05.dat}; \label{plt:sim-throughput-vs-lam-corr-fad-const-ppp-mrc-th-0dot25-p-1-N-1-fd-0dot05};
            \addplot[dashed, color=black, mark=star, mark options={scale=.8, solid}, mark repeat=1] table[x index =0, y index =4] {Results/sim-throughput-vs-lam-corr-fad-const-ppp-mrc-th-0dot25-p-1-N-1-fd-0dot1.dat}; \label{plt:sim-throughput-vs-lam-corr-fad-const-ppp-mrc-th-0dot25-p-1-N-1-fd-0dot1};
            \addplot[dashed, color=black, mark=+, mark options={scale=.8, solid}, mark repeat=1] table[x index =0, y index =4] {Results/sim-throughput-vs-lam-corr-fad-const-ppp-mrc-th-3-p-1-N-1-fd-0dot01.dat}; \label{plt:sim-throughput-vs-lam-corr-fad-const-ppp-mrc-th-3-p-1-N-1-fd-0dot01};
            \addplot[dashed, color=black, mark=o, mark options={scale=.8, solid}, mark repeat=1] table[x index =0, y index =4] {Results/sim-throughput-vs-lam-corr-fad-const-ppp-mrc-th-3-p-1-N-1-fd-0dot05.dat}; \label{plt:sim-throughput-vs-lam-corr-fad-const-ppp-mrc-th-3-p-1-N-1-fd-0dot05};
            \addplot[dashed, color=black, mark=star, mark options={scale=.8, solid}, mark repeat=1] table[x index =0, y index =4] {Results/sim-throughput-vs-lam-corr-fad-const-ppp-mrc-th-3-p-1-N-1-fd-0dot1.dat}; \label{plt:sim-throughput-vs-lam-corr-fad-const-ppp-mrc-th-3-p-1-N-1-fd-0dot1};
            \addplot[color=black, mark=diamond*, mark options={scale=1, solid}, mark repeat=1] table[x index =0, y index =1] {Results/throughput-vs-lambda-const-fading-const-ppp-th-0dot25-p-1-N-1.dat}; \label{plt:throughput-vs-lambda-const-fading-const-ppp-th-0dot25-p-1-N-1};
            \addplot[color=black, mark=diamond*, mark options={scale=1, solid}, mark repeat=1] table[x index =0, y index =1] {Results/throughput-vs-lambda-const-fading-const-ppp-th-3-p-1-N-1.dat}; \label{plt:throughput-vs-lambda-const-fading-const-ppp-th-3-p-1-N-1};
            \addplot[color=black, mark=*, mark options={scale=0.75, solid}, mark repeat=1] table[x index =0, y index =1] {Results/throughput-vs-lambda-ind-fading-const-ppp-th-0dot25-p-1-N-1.dat}; \label{plt:throughput-vs-lambda-ind-fading-const-ppp-th-0dot25-p-1-N-1};
            \addplot[color=black, mark=*, mark options={scale=0.75, solid}, mark repeat=1] table[x index =0, y index =1] {Results/throughput-vs-lambda-ind-fading-const-ppp-th-3-p-1-N-1.dat}; \label{plt:throughput-vs-lambda-ind-fading-const-ppp-th-3-p-1-N-1};

            \node at (axis cs: 1.3,0.02) {$\thr$\,=\,3};
            \node at (axis cs: 1.6,0.15) {$\thr$\,=\,0.25};
    \end{semilogyaxis}
\end{tikzpicture}

%% file: Figures/throughput-vs-efficiency.tex
\begin{tikzpicture}
    \begin{axis}[
                scale=1.0,
                xlabel={Backoff interval $\gap$},
                ylabel={Throughput $\Lambda_O$},
                xmin=1, xmax=40, ymin=0, ymax=1,
                xtick={1,10,20,30,40},
                ytick={0,0.2,0.4,0.6,0.8,1},
                yticklabels={0,0.2,0.4,0.6,0.8,1},
                grid=both,
                legend style={at={(0,0)}, anchor=south west, font=\footnotesize},
                legend cell align=left,
                legend columns=1,
                font=\footnotesize
             ]
             \addplot[thick, color=black, mark=*, mark options={scale=0.75, solid}, mark repeat=2] table[x  expr=\thisrowno{3}+1, y index =4] {Results/throughput-vs-energy-burst-ind-fad-speed-0dot1-lambda-1-p-1-thr-1-N-3.dat}; \label{plt:throughput-vs-energy-burst-ind-fad-speed-0dot1-lambda-1-p-1-thr-1-N-3};
             \addplot[thick, color=black, mark=+, mark options={scale=1.0, solid}, mark repeat=2] table[x  expr=\thisrowno{3}+1, y index =4] {Results/throughput-vs-energy-burst-ind-fad-speed-0dot1-lambda-0dot5-p-1-thr-0dot1-N-3.dat}; \label{plt:throughput-vs-energy-burst-ind-fad-speed-0dot1-lambda-0dot5-p-1-thr-0dot1-N-3};
             \addplot[thick, color=black, mark=star, mark options={scale=1.0, solid}, mark repeat=2] table[x  expr=\thisrowno{3}+1, y index =4] {Results/throughput-vs-energy-burst-ind-fad-speed-0dot1-lambda-0dot75-p-1-thr-0dot25-N-3.dat}; \label{plt:throughput-vs-energy-burst-ind-fad-speed-0dot1-lambda-0dot75-p-1-thr-0dot25-N-3};
             \addplot[thick, dashed, color=black, mark=*, mark options={scale=0.75, solid}, mark repeat=2] table[x  expr=\thisrowno{3}+1, y index =4] {Results/throughput-vs-energy-burst-ind-fad-speed-0dot1-lambda-1-p-1-thr-1-N-1.dat}; \label{plt:throughput-vs-energy-burst-ind-fad-speed-0dot1-lambda-1-p-1-thr-1-N-1};
             \addplot[thick, dashed, color=black, mark=+, mark options={scale=1.0, solid}, mark repeat=2] table[x  expr=\thisrowno{3}+1, y index =4] {Results/throughput-vs-energy-burst-ind-fad-speed-0dot1-lambda-0dot5-p-1-thr-0dot1-N-1.dat}; \label{plt:throughput-vs-energy-burst-ind-fad-speed-0dot1-lambda-0dot5-p-1-thr-0dot1-N-1};
             \addplot[thick, dashed, color=black, mark=star, mark options={scale=1.0, solid}, mark repeat=2] table[x  expr=\thisrowno{3}+1, y index =4] {Results/throughput-vs-energy-burst-ind-fad-speed-0dot1-lambda-0dot75-p-1-thr-0dot25-N-1.dat}; \label{plt:throughput-vs-energy-burst-ind-fad-speed-0dot1-lambda-0dot75-p-1-thr-0dot25-N-1};
    \end{axis}
    \begin{axis}[
                scale=1.0,
                ylabel={Energy efficiency $\efficiency$},
                xmin=1, xmax=40, ymin=0, ymax=1,
                axis y line*=right,
                ytick={0,0.2,0.4,0.6,0.8,1},
                yticklabels={0,0.2,0.4,0.6,0.8,1},
                grid=both,
                legend style={at={(0,0)}, anchor=south west, font=\footnotesize},
                legend cell align=left,
                legend columns=1,
                font=\footnotesize
             ]
             \addplot[color=black, mark=*, mark options={scale=0.75, solid}, mark repeat=2] table[x  expr=\thisrowno{3}+1, y index =6] {Results/throughput-vs-energy-burst-ind-fad-speed-0dot1-lambda-1-p-1-thr-1-N-3.dat}; \label{plt:energy-efficiency-burst-ind-fad-speed-0dot1-lambda-1-p-1-thr-1-N-3};
             \addplot[color=black, mark=+, mark options={scale=1.0, solid}, mark repeat=2] table[x  expr=\thisrowno{3}+1, y index =6] {Results/throughput-vs-energy-burst-ind-fad-speed-0dot1-lambda-0dot5-p-1-thr-0dot1-N-3.dat}; \label{plt:energy-efficiency-burst-ind-fad-speed-0dot1-lambda-0dot5-p-1-thr-0dot1-N-3};
             \addplot[color=black, mark=star, mark options={scale=1.0, solid}, mark repeat=2] table[x  expr=\thisrowno{3}+1, y index =6] {Results/throughput-vs-energy-burst-ind-fad-speed-0dot1-lambda-0dot75-p-1-thr-0dot25-N-3.dat}; \label{plt:energy-efficiency-burst-ind-fad-speed-0dot1-lambda-0dot75-p-1-thr-0dot25-N-3};
             \addplot[dashed, color=black, mark=*, mark options={scale=0.75, solid}, mark repeat=2] table[x  expr=\thisrowno{3}+1, y index =6] {Results/throughput-vs-energy-burst-ind-fad-speed-0dot1-lambda-1-p-1-thr-1-N-1.dat}; \label{plt:energy-efficiency-burst-ind-fad-speed-0dot1-lambda-1-p-1-thr-1-N-1};
             \addplot[dashed, color=black, mark=+, mark options={scale=1.0, solid}, mark repeat=2] table[x  expr=\thisrowno{3}+1, y index =6] {Results/throughput-vs-energy-burst-ind-fad-speed-0dot1-lambda-0dot5-p-1-thr-0dot1-N-1.dat}; \label{plt:energy-efficiency-burst-ind-fad-speed-0dot1-lambda-0dot5-p-1-thr-0dot1-N-1};
             \addplot[dashed, color=black, mark=star, mark options={scale=1.0, solid}, mark repeat=2] table[x  expr=\thisrowno{3}+1, y index =6] {Results/throughput-vs-energy-burst-ind-fad-speed-0dot1-lambda-0dot75-p-1-thr-0dot25-N-1.dat}; \label{plt:energy-efficiency-burst-ind-fad-speed-0dot1-lambda-0dot75-p-1-thr-0dot25-N-1};
    \end{axis}
\end{tikzpicture}